\documentclass[12pt]{article}

\usepackage{times}

\usepackage{amsmath} 
\usepackage{amssymb}
\usepackage{bm}
\usepackage[ruled,vlined]{algorithm2e} 
\usepackage{algorithmic}
\usepackage{graphicx}
\usepackage{float}
\usepackage[toc,page]{appendix}
\usepackage{hyperref}
\usepackage{cleveref}
\usepackage{geometry}
\usepackage{nameref}
\usepackage{xcolor} 
\usepackage{caption}
\usepackage{subcaption}
\usepackage{comment}
\usepackage{lineno}

\newcommand{\Regc}{\text{\textbf{b}}} 
\newcommand{\regc}{\text{b}} 
\newcommand{\PrecisionN}{\mathbf{\Theta}}
\newcommand{\SampleCov}{\text{\textbf{S}}}
\newcommand{\argmax}{\mathop{\mathrm{argmax}}} 
\newcommand{\noind}{\not\!\perp\!\!\!\perp}
\newcommand{\indep}{\!\perp\!\!\!\perp}
\renewcommand*{\mathbf}[1]{\ifmmode\bm{#1}\else\textbf{#1}\fi} 

\title{New estimation approaches for graphical models with elastic net penalty}

\author
{Davide Bernardini,$^{1}$ Sandra Paterlini,$^{1}$ Emanuele Taufer$^{1}$\\
\\
\normalsize{$^{1}$Department of Economics and Management, University of Trento,}\\
}

\date{}

\begin{document}

\maketitle

\begin{abstract}
In the context of undirected Gaussian graphical models, we introduce three estimators based on elastic net penalty for the underlying dependence graph.  Our goal is to estimate the sparse precision matrix, from which to retrieve both the underlying conditional dependence graph and the partial correlation graph. The first estimator is derived from the direct penalization of the precision matrix in the likelihood function, while  the second from using conditional penalized regressions to estimate the precision matrix. Finally, the third estimator relies on a 2-stages procedure that estimates the edge set first and then the precision matrix elements. Through simulations we investigate the performances of the proposed methods on a large set of well-known network structures. Empirical results on simulated data show that the 2-stages procedure outperforms all other estimators  both w.r.t. estimating the sparsity pattern in the graph and the edges’ weights. Finally, using real-world data on US economic sectors, we estimate dependencies and show the impact of Covid-19 pandemic on the network strength.
\end{abstract}


\baselineskip24pt


\newpage
\section{Introduction}
\noindent
Let $\mathbf{X} = [X_1,X_2,...,X_p]^\top$ be a $p$-dimensional multivariate Gaussian  random vector, denoted with $\mathcal{N}_p(\mathbf{\mu},\mathbf{\Sigma})$, where $\mathbf{\mu}$ is the mean vector and $\mathbf{\Sigma}$ is the covariance matrix. The matrix $\PrecisionN = \mathbf{\Sigma}^{-1}$ is often called precision (or concentration) matrix. In this paper, we consider the problem of estimating a sparse $\PrecisionN$ in a multivariate Gaussian framework by introducing three estimation techniques based on the elastic net penalty \cite{ElNet05}.

\noindent
Graphical models are useful for representing a set of random variables and their conditional dependence structure. A graphical model is made by two elements: a graph $\mathcal{G}(\text{\textbf{N}}, \text{\textbf{E}})$ and a joint distribution $\mathbf{f}$. The set of nodes $\text{\textbf{N}} = \{1,2,...,p\}$ of the graph represents random variables over which the joint distribution is defined. Instead, the edge set \textbf{E} represents the pairs of variables which are conditionally dependent given the remaining ones. Thus, a pair $(i,j) \in \text{\textbf{E}}$, with $i,j = 1,...,p \wedge i\neq j$, if and only if $X_i \noind X_j|X_C$ where $C=\{k \in \text{\textbf{N}}|k\neq i,j\}$ and $\indep$ indicates independence between two random variables. For a detailed description of graphical models, see for example Lauritzen \cite{GM1996} or Koller and Friedman \cite{GM2009}. Here, we consider undirected Gaussian graphical models where the joint distribution of random variables is multivariate normal. In this special case, the dependence structure among each component in the multivariate distribution is obtainable through  $\PrecisionN$, see \cite{GM1996} for example; if $\theta_{ij}$ is the ($i,j$) element of $\PrecisionN$, then $\theta_{ij}=0$ if and only if $X_i \indep X_j|X_C$.

\noindent
Estimation of $\PrecisionN$ in the Gaussian setting may serve other purposes. A sparse estimate $\hat{\PrecisionN}$ of a sparse $\PrecisionN$ allows us not only to reconstruct the conditional dependence graph, but also the partial correlation graph. The latter can be useful to model how shocks in the variables propagate in the whole system,  as discussed for example by Anufriev and Panchenko \cite{anuf15}. 

\noindent
In this article, we introduce and test through simulations three estimators for the precision matrix $\PrecisionN$. These methods rely on penalization techniques to produce a sparse estimate. Penalized estimation methods are used nowadays to estimate sparse and typically more interpretable models. One could see them as noise filter techniques able to retrieve the relevant information from data. In the context of linear regression, Tibshirani \cite{Tib96} proposed the \emph{lasso} regression to achieve sparsity in the estimated vector of regressors' coefficients. By adding a penalty equal to the $\ell_1$-norm of the coefficients' vector to the standard OLS problem, \emph{lasso} allows to automatically perform model selection and estimation in a single step. \emph{Lasso} can then deal with  high-dimensional situations, where the number of parameters to estimate is larger than the sample size of data available. Beside regression, the \emph{lasso} or a similar approach could then also be considered to estimate the sparse dependence graph of a Gaussian graphical model. For example, Meinshausen and B\"uhlmann \cite{Mein06} used \emph{lasso} regressions to estimate the edge set of the graph. Banerjee et al. \cite{Bane08} and Friedman et al. \cite{glasso08} used an element-wise $\ell_1$-norm to penalize the precision matrix in the log-likelihood function of the multivariate Gaussian distribution. Differently from the approach of Meinshausen and B\"uhlmann \cite{Mein06}, Banerjee et al. \cite{Bane08} and Friedman et al. \cite{glasso08} are able not only to perform edge set selection, but also to produce a sparse estimate of $\PrecisionN$.

\noindent
Starting from the original article about \emph{lasso} in linear regression \cite{Tib96}, several modifications have been proposed in the literature. Among them, one of the most successful and widely used is  the elastic net penalty, proposed by Zou and Hastie \cite{ElNet05}. In its original formulation, elastic net penalty consists of adding an additional $\ell_2$-norm to the $\ell_1$-norm penalty in the objective function of the OLS regression problem. According to \cite{ElNet05}, elastic net is then  capable to address some of the limitations of the \emph{lasso} approach, such as the seemingly random selection among highly correlated variables or, in the low-dimensional case, the worse prediction performance with respect to ridge regression in presence of high correlation. Contrary to the wide usage of \emph{lasso} type penalty in the estimation of conditional dependence graphs, the elastic net penalty has so far received limited attention. To the best of our knowledge only few recent works use an elastic net type penalty for graphical models, see Cucuringu et al. \cite{Cucu11}, Ryali et al. \cite{Ryal12} and Liu et al. \cite{Liu16}. Cucuringu et al. \cite{Cucu11} relies on a neighbors selection approach similar to \cite{Mein06}, which is  related to the second and third estimator we propose in this paper (see section \ref{CRgelnetdesc} and \ref{2Sgelnetdesc}). Still, our goal is to produce a sparse estimate of $\PrecisionN$, not only to estimate the dependence graph. Ryali et al. \cite{Ryal12} rely on a maximum likelihood approach, which is connected to our first estimation method (see section \ref{gelnetdesc}). However, compared to Ryali et al.\cite{Ryal12}, we derive an optimization procedure inspired by \cite{glasso08}. 

\noindent
In this paper, after introducing the three proposed estimators of $\PrecisionN$,  we propose an extensive comparison among them and \emph{glasso} \cite{glasso08} on well-known network structures.  In fact, we test them on a large spectrum of graph's topologies, such as scale-free, small-world and core-periphery networks. Such topologies can be observed in the real-world, such as the scale-free property being observed in the network of Web pages \cite{scalefree99}, while small-world properties are detected in social networks and neural networks \cite{smallworld98}. Empirical results on simulated data allow to compare the validity of the proposed approaches. Then, we focus on a real-world application directed to estimate dependencies across US sectors and evaluate the impact of Covid-19 pandemic on the network strength.

\noindent
This paper is structured as follows. Section \ref{sec:description} describes the three estimation approaches we propose. In section \ref{sec:setup} we report the simulations' settings, while in section \ref{sec:results} we discuss the empirical results. In section \ref{sec:empana}, we present an application using real-world data. To conclude we discuss briefly our main findings in section \ref{sec:conclusions}.
\newpage
\section{Elastic Net Estimators} \label{sec:description}
\noindent
 We propose here three different estimation methods that rely on the elastic net penalty for the estimation of sparse precision matrices. 
 
 \noindent
 The first method results in  an estimate of matrix $\PrecisionN$ through a penalized log-likelihood optimization similar to the procedure used in the graphical lasso (\emph{glasso}) \cite{glasso08}. The penalty used is a combination of element-wise $\ell_1$-norm and $\ell_2$-norm of a matrix. Here, we call this approach graphical elastic net, with acronym \emph{gelnet}.
 
 \noindent
 The second method is inspired by the approach of Yuan \cite{Yuan10}  for high-dimensional inverse covariance matrix estimation. It uses conditional regressions with elastic net penalty to produce an asymmetric estimate of $\PrecisionN$. We then use a procedure to derive a symmetric estimate starting from the first raw and asymmetric one. We call this second approach conditional regressions graphical elastic net, with acronym \emph{CR-gelnet}.
 
 \noindent
 The third method relies on  a combination of the first two. At first, it uses conditional regressions with elastic net penalty to estimate the sparsity structure, that is the positions of zeroes, of the matrix $\PrecisionN$. This is similar to the neighbourhood selection of Meinshausen and B\"uhlmann \cite{Mein06}. Then, the method proceeds with a constrained maximum likelihood estimation of $\PrecisionN$, where the constraints are the zero off-diagonal elements found in the previous step. Because of this two stages procedure, we call this final approach two stages graphical elastic net, with acronym \emph{2S-gelnet}.
 
\vspace{0.5cm}
\subsection{Graphical elastic net - [\emph{gelnet}]} \label{gelnetdesc}
\noindent
The first method follows closely the approach proposed by Friedman et al. \cite{glasso08}, thus it is a penalized log-likelihood optimization problem. We suggest a direct modification of the original algorithm by \cite{glasso08}, taking into account an additional part in the penalty term. The new augmented penalty is  a convex linear combination of two penalties defined respectively as element-wise $\ell_1$ and $\ell_2$ norms of precision matrix $\PrecisionN$. This augmented penalty leads to the following convex optimization problem:
\begin{equation}
	\hat{\PrecisionN} = \argmax_{\PrecisionN} \Big{\{} \text{log}(\text{det}(\PrecisionN)) - \text{trace}(\SampleCov\PrecisionN) - \alpha\lambda||\PrecisionN||_1 - (1-\alpha)\lambda||\PrecisionN||_2^2 \Big{\}}
\end{equation}
where:
\begin{itemize}
    \item $||\PrecisionN||_1 = \sum_{i=1}^{p}\sum_{j=1;j\neq i}^{p}|\theta_{ij}|$ is the element-wise $\ell_1$-norm of matrix $\PrecisionN$
    \item $||\PrecisionN||_2 = \sqrt{\sum_{i=1}^{p}\sum_{j=1;j\neq i}^{p}(\theta_{ij})^2}$ is the element-wise $\ell_2$-norm of matrix $\PrecisionN$
    \item $\SampleCov$ is the sample covariance matrix of data
    \item $\alpha \in [0,1]$ is the first penalty parameter. This is used to control the balance between $\ell_1$-norm and $\ell_2$-norm penalties. Note that when $\alpha=1$ we have the graphical lasso optimization problem.
    \item $\lambda > 0$ is the second penalty parameter and it is used to control the strength of the penalty.
\end{itemize}
This is the penalized log-likelihood of the Gaussian multivariate distribution $\mathcal{N}_p(\mathbf{\mu},\mathbf{\Sigma)}$ already partially maximized \footnote{Partially maximized because we use the estimate of $\mathbf{\mu}$ to compute $\SampleCov$} with respect to mean vector $\mathbf{\mu}$ and ignoring the constant term. 
We exclude diagonal elements from the element-wise matrix norms as we want to shrink and induce sparsity in the off-diagonal elements only.
\newline
We derive the penalized log-likelihood with respect to $\PrecisionN$, similarly to \cite{glasso08} using rules from \cite{bookCO2004}.
So we obtained the following optimal conditions:
\begin{equation}
	\PrecisionN^{-1} - \SampleCov - \alpha\lambda\mathbf{\Gamma} - 2(1-\alpha)\lambda\PrecisionN = \mathbf{0}
\end{equation}
where $\mathbf{\Gamma}$ is the subgradient matrix whose elements are $\gamma_{ij} = 1$ if $\theta_{ij} > 0$, $\gamma_{ij} = [-1,1]$ if $\theta_{ij} = 0$ and $\gamma_{ij} = -1$ if $\theta_{ij} < 0$.
\newline
The block coordinate descent algorithm proposed for graphical lasso \cite{glasso08} can be adapted to our new optimization problem with elastic net penalty. This algorithm is similar to a coordinate descent algorithm with the difference that it updates a block of variables (here the elements of $j$-th row/column) at a time leaving the rest of them fixed.
\newline
Let $\text{\textbf{W}}$ the estimate of $\mathbf{\Sigma}$ and consider the following notation for the partitioning of a generic matrix \textbf{A}. Let \textbf{A}$_{11}$ represents matrix \textbf{A} with row and column $k$-th removed, \textbf{a}$_{12}$ the $k$-th column of \textbf{A} with $k$-th row removed, \textbf{a}$_{21}$ the $k$-th row with $k$-th column removed (equal to \textbf{a}$_{12}$ if \textbf{A} is symmetric) and a$_{22}$ is the element of \textbf{A} in the $k$-th row and $k$-th column.
\begin{equation} \label{matsub}
\centering
\text{\textbf{A}} = \begin{bmatrix} 
\text{\textbf{A}}_{11} & \text{\textbf{a}}_{12} \\
\text{\textbf{a}}_{21} & \text{a}_{22} 
\end{bmatrix} 
\end{equation}
The block coordinate method cycles through each $k$-th row/column solving for each $k$ the following subproblem, while keeping the other values fixed:
\begin{equation}
	\text{\textbf{w}}_{12} - \text{\textbf{s}}_{12} - \alpha\lambda\mathbf{\gamma}_{12} - 2(1-\alpha)\lambda\mathbf{\theta}_{12} = \mathbf{0}
\end{equation}
and we set $w_{kk} = s_{kk}$, not penalizing the diagonal.
\newline
Then, using the formula for block-partitioned matrix \cite{glasso08}, we have that:
\begin{equation}
\begin{bmatrix} 
\text{\textbf{W}}_{11} & \text{\textbf{w}}_{12} \\
\text{\textbf{w}}_{21} & \text{w}_{22} 
\end{bmatrix} = \begin{bmatrix} 
\boldsymbol{\Theta}_{11} - \Big{(}\frac{\boldsymbol{\theta}_{12}\boldsymbol{\theta}_{21}}{\theta_{22}}\Big{)}^{-1} & -\text{\textbf{W}}_{11} \frac{\boldsymbol{\theta}_{12}}{\theta_{22}}\\
\Big{(}-\text{\textbf{W}}_{11} \frac{\boldsymbol{\theta}_{12}}{\theta_{22}}\Big{)}^T & \frac{1}{\theta_{22}}-\frac{\boldsymbol{\theta}_{21}\text{\textbf{W}}_{11}\boldsymbol{\theta}_{12}}{\theta_{22}^2}
\end{bmatrix}
\end{equation}
Substituting $\text{\textbf{w}}_{12}$ with $-\text{\textbf{W}}_{11} \frac{\mathbf{\theta}_{12}}{\theta_{22}}$ and setting $\Regc=-\frac{\mathbf{\theta}_{12}}{\theta_{22}}$ we can rewrite the optimal condition as follows:
\begin{equation}
	\text{\textbf{W}}_{11}\Regc - \text{\textbf{s}}_{12} + \alpha\lambda\mathbf{\gamma}_{12} + 2(1-\alpha)\lambda\theta_{22}\Regc = \mathbf{0}
\end{equation}
\newline
These optimal conditions resemble closely the normal equations of the linear regression with elastic net penalty term. In fact, it is possible to use the coordinate descent approach \cite{Fried10} to find the optimal solution of \textbf{b}. The optimal updates for each block of variables can be derived as follows:
\begin{equation}
 \sum_{t \neq j}^{p-1}\text{\textbf{W}}_{11,jt}\regc_t + \text{\textbf{W}}_{11,jj}\regc_j - \text{\textbf{s}}_{12,j} + 2(1-\alpha)\lambda\theta_{22}\regc_j + 
\setlength\arraycolsep{1pt}
\left\{
\begin{array}{cl}
-\alpha\lambda & \text{ if } \regc_j < 0  \\
\big{[}-\alpha\lambda, \alpha\lambda\big{]} & \text{ if } \regc_j = 0 \\
\alpha\lambda & \text{ if } \regc_j > 0
\end{array}
\right.
\end{equation}
Setting $c_j = \sum_{t \neq j}^{p-1}\text{\textbf{W}}_{11,jt}\regc_t - \text{\textbf{s}}_{12,j}$ and $h_j = \text{\textbf{W}}_{11,jj}$, we have for each variable in the current block:
\begin{equation}
\left\{
\begin{array}{cl}
c_j + h_j\regc_j + 2(1-\alpha)\lambda\theta_{22}\regc_j -\alpha\lambda & \text{ if } \regc_j < 0  \\
\big{[}c_j-\alpha\lambda, c_j+\alpha\lambda\big{]} & \text{ if } \regc_j = 0 \\
c_j + h_j\regc_j + 2(1-\alpha)\lambda\theta_{22}\regc_j + \alpha\lambda & \text{ if } \regc_j > 0
\end{array}
\right.
\end{equation}
Therefore the optimal updates are:
\begin{itemize}
	\item $\regc_j^* = \frac{-c_j+\alpha\lambda}{h_j+ 2(1-\alpha)\lambda\theta_{22}}$ if $\regc_j < 0$ thus $-c_j+\alpha\lambda < 0$ $\Rightarrow$ $c_j > \alpha\lambda$
	\item $\regc_j^* = 0$ with $[c_j-\alpha\lambda, c_j+\alpha\lambda]$ containing 0, thus $\left\{
														\begin{array}{cl}
															c_j - \alpha\lambda \leq 0 \\
															c_j + \alpha\lambda \geq 0
														\end{array}
																					\right.$ $\Rightarrow$ $-\alpha\lambda \leq c_j \leq \alpha\lambda$
	\item $\regc_j^* = \frac{-c_j-\alpha\lambda}{h_j+ 2(1-\alpha)\lambda\theta_{22}}$ if $\regc_j > 0$ thus $-c_j-\alpha\lambda > 0$ $\Rightarrow$ $c_j < -\alpha\lambda$
\end{itemize}
or using soft-thresholding operator:
\begin{equation}
\begin{split}
 & \text{Soft}\Big{(}c_j,\alpha\lambda\Big{)} := \text{sign}(c_j)\Big{(}|c_j|-\alpha\lambda\Big{)}_+ \\
 \regc_j^* &= \frac{-\text{Soft}(c_j,\alpha\lambda)}{h_j+2(1-\alpha)\lambda\theta_{22}}
\end{split}
\end{equation}
Once the coordinate descent algorithm has converged to the optimal values $\Regc^*$, we can update $\text{\textbf{w}}_{12}$:
\begin{equation}
\text{\textbf{w}}_{12} = \text{\textbf{W}}_{11}\Regc^*
\end{equation}
then $\theta_{22}$ from:
\begin{equation}
\theta_{22} = \frac{1}{\text{w}_{22}-\Regc^*\text{\textbf{w}}_{12}}
\end{equation}
and finally $\mathbf{\theta}_{12}$ from:
\begin{equation}
\mathbf{\theta}_{12} = -\Regc^*\theta_{22}
\end{equation}
The difference here between the original graphical lasso algorithm is that at each block-update $\theta_{22}$ and $\mathbf{\theta}_{12}$ are also updated.
\newline
The block coordinate algorithm then proceeds with the updates of the next block, which is the next row/column of the matrix \textbf{W}. After all $p$ rows/columns are updated, numerical convergence is checked. This is performed by checking if the biggest difference, in absolute value, among the elements of two full subsequent updates of \textbf{W} is less than a specific threshold $\delta$. After the algorithm converges, the final updated version of $\PrecisionN$ is the estimate $\hat{\PrecisionN}$ of the precision matrix and the final updated version of \textbf{W} is the estimate of $\mathbf{\Sigma}$.
\newline
Pseudo-code is reported in \hyperref[alg1]{\textbf{Algorithm 1}}:
\begin{algorithm} \label{alg1}
\caption{\emph{gelnet}}
\label{gelnet}
\begin{algorithmic}
\STATE Set \textbf{W} = \textbf{S} and $\mathbf{\Theta}$ = \textbf{S}$^{-1}$
\STATE Set  a threshold value $\delta$
\STATE Set Convergence = FALSE
\WHILE{Convergence==FALSE}
\STATE \textbf{W}$_{\text{old}}$ = \textbf{W}
\FOR{k=1,..,p}
\STATE 1) Subdivide \textbf{W}, \textbf{S} and $\mathbf{\Theta}$ into blocks as in formula \ref{matsub}
\STATE 2) Evaluate using a coordinate descent algorithm the optimal values $\Regc^*$ of $\Regc$. In the first update of \textbf{W} set randomly their initial values, while in the following updates of \textbf{W} use their previous estimates.
\STATE 3) Use $\Regc^*$ to update the values of \textbf{w}$_{12}$, $\theta_{22}$ and $\mathbf{\theta}_{12}$
\STATE 4) Store $\Regc^*$, these will be the next starting point for the coordinate descent for this block of variables in the next update of the matrices
\ENDFOR
\IF{$\max_{i,j}$(abs(\textbf{W}$_{\text{old},ij}$ - \textbf{W}$_{ij}$)) $<$ $\delta$}
\STATE Convergence=TRUE
\ENDIF
\ENDWHILE

\end{algorithmic}
\end{algorithm}
\newline

\subsection{Conditional regressions graphical elastic net - [\emph{CR-gelnet}]}
\label{CRgelnetdesc}
\noindent
The conditional regression method we propose follows  the approach of Yuan \cite{Yuan10} and Bogdan et al. \cite{gslope18}. The core idea of this approach is to fit an elastic net penalized regression for each component of a multivariate Gaussian random vector. That is, the $i$-th element of the random vector becomes the dependent variable while the other elements are the independent variables. These coefficients of linear regressions are proportional to the elements of $\PrecisionN$. Thus, given a $p$-dimensional random variable, the estimated coefficients of each penalized regression fitted are used to produce an estimate of $\PrecisionN$. Since this estimate is not symmetric, an additional procedure must be employed to produce a symmetric estimate of $\PrecisionN$. In the following part of this subsection, we describe in more details the whole procedure.
\newline
Let $\mathbf{X}$ the $p$-dimensional random vector with Gaussian distribution. The conditional distribution of $i$-th component $X_i$, given the remaining components $\mathbf{X}_{-i}$ can be expressed as linear relation \cite{Yuan10}:
\begin{equation}
	X_i|\mathbf{X}_{-i} = \text{a}_i +  \mathbf{X}_{-i}^\top\Regc_i + \epsilon_i
\end{equation}
with $\text{a}_i = \mu_i - \mathbf{\Sigma}_{i,-i}\mathbf{\Sigma}_{-i,-i}^{-1}\mathbf{\mu}_{-i}$ and $\Regc_i = \mathbf{\Sigma}_{-i,-i}^{-1}\mathbf{\Sigma}_{-i,i}$
If $\mathbf{X} \sim \mathcal{N}(\mathbf{\mu},\mathbf{\Sigma})$, then \cite{Yuan10} \cite{bookMA1979}:
\begin{itemize}
	\item random error term $\epsilon_i \sim \mathcal{N}(0,\mathbf{\Sigma}_{ii}-\mathbf{\Sigma}_{i,-i}\mathbf{\Sigma}_{-i,-i}^{-1}\mathbf{\Sigma}_{-i,i})$
	\item the conditional distribution of $X_i|\mathbf{X}_{-i} \sim \mathcal{N}(\mu_i + \mathbf{\Sigma}_{i,-i}\mathbf{\Sigma}_{-i,-i}^{-1}(\mathbf{X}_{-i}-\mathbf{\mu}_{-i}), \mathbf{\Sigma}_{ii}-\mathbf{\Sigma}_{i,-i}\mathbf{\Sigma}_{-i,-i}^{-1}\mathbf{\Sigma}_{-i,i})$
\end{itemize}
where $\mathbf{\Sigma}_{-i,-i}$ refers to the covariance matrix $\mathbf{\Sigma}$ without the $i$-th row and column, $\mathbf{\Sigma}_{i,i}$ to the element of $\mathbf{\Sigma}$ in the $i$-th column and row, $\mathbf{\Sigma}_{i,-i}$ to the $i$-th row of $\mathbf{\Sigma}$ but without the $i$-th element, $\mathbf{\Sigma}_{-i,i}$ to 
the $i$-th column of $\mathbf{\Sigma}$ but without the $i$-th element.
\newline
Setting z$_i$ = $(\boldsymbol{\Sigma}_{i,i}-\boldsymbol{\Sigma}_{i,-i}\boldsymbol{\Sigma}_{-i,-i}^{-1}\boldsymbol{\Sigma}_{-i,i})^{-1}$ and using the rule for the inverse of a matrix subdivided in blocks, as in \cite{Yuan10} for example, we have that:
\begin{equation}
\begin{bmatrix} 
\boldsymbol{\Sigma}_{i,i} &  \boldsymbol{\Sigma}_{i,-i}\\
\boldsymbol{\Sigma}_{-i,i} & \boldsymbol{\Sigma}_{-i,-i}
\end{bmatrix}^{-1} =
\end{equation}

\begin{equation}
= \begin{bmatrix} 
\text{z}_i & -\text{z}_i\boldsymbol{\Sigma}_{i,-i}\boldsymbol{\Sigma}_{-i,i}^{-1} \\
-\boldsymbol{\Sigma}_{-i,i}^{-1}\boldsymbol{\Sigma}_{-i,i}\text{z}_i & \boldsymbol{\Sigma}_{-i,-i}^{-1}+\boldsymbol{\Sigma}_{-i,-i}^{-1}\boldsymbol{\Sigma}_{-i,i}\text{z}_i\boldsymbol{\Sigma}_{i,-i}\boldsymbol{\Sigma}_{-i,-i}^{-1}
\end{bmatrix} =
\end{equation}

\begin{equation}
= \begin{bmatrix}
\boldsymbol{\PrecisionN}_{i,i} & \boldsymbol{\PrecisionN}_{i,-i} \\
\boldsymbol{\PrecisionN}_{-i,i} & \boldsymbol{\PrecisionN}_{-i,-i}
\end{bmatrix}
\end{equation}
It is clear from the formula above that $\PrecisionN_{i,i} = \frac{1}{\text{Var}(\epsilon_i)}$ and $\PrecisionN_{-i,i} = -\frac{\Regc_i}{\text{Var}(\epsilon_i)}$. Therefore the sparsity pattern in the regression coefficients $\Regc_i$ corresponds to the sparsity pattern in the off-diagonal elements of the $i$-th column of $\PrecisionN$. Thus ideally the matrix 
$\PrecisionN$ can be reconstructed by properly rescaling the regression coefficients $\Regc_i$. This is the justification behind this conditional regression approach for the estimation of a sparse $\PrecisionN$.
\newline
The estimation of the full matrix $\PrecisionN$ is done by the estimation of $p$ penalized regressions, one for each component of multivariate random variable on the remaining $p-1$ components. Yuan \cite{Yuan10} used the Dantzig selector for the estimation of $\Regc_i$, but here we use $p$ elastic net regressions all having the same parameters $\alpha$ and $\lambda$. Both estimation techniques can produce sparse estimates $\hat{\Regc}_i$ of $\Regc_i$, thus we can obtain a sparse estimate $\hat{\PrecisionN}$ of $\PrecisionN$.
\newline
Since this approach doesn't guarantee that the estimated matrix $\hat{\PrecisionN}$ is symmetric, an additional procedure is needed to make this estimate symmetric. Ideally a good procedure here should be able to retain sparsity. We propose and use two simple and computationally cheap procedures:
\begin{itemize}
	\item \textbf{L2 method (L2)}: we replace the elements $\hat{\theta}_{ij}$ and $\hat{\theta}_{ji}$ of the initial asymmetric estimated matrix $\hat{\PrecisionN}$ with their average. This is analogous to minimize the Frobenius distance (element-wise $\ell_2$-norm) between the original asymmetric matrix and a symmetric matrix as in \cite{gslope18}. Note that this rule retains sparsity when both elements ($i,j$) and ($j,i$) are zero.
	\item \textbf{Minimum element method (MinEl)}: we substitute the elements $\hat{\theta}_{ij}$ and $\hat{\theta}_{ji}$ with the minimum element, in absolute value, between elements ($i,j$) and ($j,i$) of the asymmetric estimate $\hat{\PrecisionN}$, as proposed by Cai et al. \cite{Cai11}. Note that this rule retains sparsity when at least one element between ($i,j$) and ($j,i$) is zero.
\end{itemize}
The conditional regression approach for estimating the matrix $\PrecisionN$ can be summarized as in \hyperref[alg2]{\textbf{Algorithm 2}}:
\begin{algorithm} \label{alg2}
\caption{\emph{CR-gelnet}}
\begin{algorithmic}
\STATE Set $\hat{\PrecisionN}$ as \textbf{I}$_{pxp}$ ($p$ by $p$ identity matrix)
\FOR{i=1,..,p}
\STATE 1) Fit an elastic net penalized regression using coordinate descent procedure in order to estimate the coefficients' vector $\Regc_i$ of $\text{a}_i +  \mathbf{X}_{-i}^T\Regc_i + \epsilon_i$
\STATE 2) Use residuals of the regression to estimate Var($\epsilon_i$)
\STATE 3) Given the estimated values $\hat{\Regc}_i$ and $\hat{\text{Var}}$($\epsilon_i$), update the values of $\hat{\PrecisionN}_{i,i}$ and $\hat{\PrecisionN}_{-i,i}$
\ENDFOR
\STATE Use MinEl or L2 method to make the estimated $\hat{\PrecisionN}$ symmetric.
\end{algorithmic}
\end{algorithm}

\subsection{Two stages graphical elastic net - [\emph{2S-gelnet}]}
\label{2Sgelnetdesc}
\noindent
As pointed out at the beginning of the section, this last approach can be seen as a mix of the two methods already discussed. It consists of 2 stages. Firstly we use elastic net penalized conditional regressions to estimate the sparsity pattern in the matrix $\PrecisionN$. Secondly we use the estimated positions of zeros to do a constrained log-likelihood estimation of $\PrecisionN$, given that the distribution of data is a multivariate Gaussian.

\noindent
Meinshausen and B\"uhlmann \cite{Mein06} suggested that the edge set \textbf{E} of the conditional dependence graph of an undirected Gaussian graphical model can be estimated by using penalized regressions able to induce sparsity in the estimated coefficients. Their approach is usually called neighbourhood selection and its aim is only to estimate the edge set \textbf{E} of a Gaussian graphical model, but not the estimation of the precision matrix. They used conditional \emph{lasso} regressions to reconstruct a sparse graph. A zero regression coefficient implies zero partial correlation and thus the absence of the related edge in the graph, at least in the multivariate Gaussian situation. For our estimation problem, it means a zero in the precision matrix, thus a zero coefficient constraint in the log-likelihood optimization problem. We can easily extend the approach of Meinshausen and B\"uhlmann \cite{Mein06} using conditional elastic net regressions instead of \emph{lasso} regressions to estimate the edge set of the graph of a Gaussian graphical model.
\newline
Let \textbf{X} be the $n$ by $p$ matrix of observations, we fit $p$ conditional elastic net penalized regressions of each component over the remaining ones:
\begin{equation}
  \hat{\text{\textbf{b}}}_i = \text{argmin}_{\text{\textbf{b}}_i}\Big{\{}||\text{\textbf{X}}_i-\text{a}_i-\text{\textbf{X}}_{-i}\text{\textbf{b}}_i||_2^2+\lambda[\alpha||\text{\textbf{b}}_i||_1+(1-\alpha)||\text{\textbf{b}}_i||_2^2]\Big{\}}
\end{equation}
where \textbf{X}$_i$ is the $i$-th column of \textbf{X} and \textbf{X}$_{-i}$ is the matrix \textbf{X} without $i$-th column.
\noindent
This approach could lead to different conclusions about the inclusion of the edge $(j,i)$ (undirected edge between $j$ and $i$) because the estimated regression coefficient $\hat{\regc}_{ji}$ ($j$-th element of $\hat{\Regc}_i$) is not necessary zero when $\hat{\regc}_{ij}$ ($i$-th element of $\hat{\Regc}_j$) is zero, or vice versa. Meinshausen and B\"uhlmann \cite{Mein06} suggested two possible rules to deal with this situation:
\begin{itemize}
    \item AND rule: edge $(i,j) \in \hat{\text{\textbf{E}}}$ if $\hat{\regc}_{ij}\neq0 \wedge \hat{\regc}_{ji}\neq0$
    \item OR rule: edge $(i,j) \in \hat{\text{\textbf{E}}}$ if $\hat{\regc}_{ij}\neq0 \vee \hat{\regc}_{ji}\neq0$
\end{itemize}
where $\hat{\text{\textbf{E}}}$ represent the estimated edge set of the underlying undirected Gaussian graphical model. We follow these two rules to set the zero positions in the precision matrix $\PrecisionN$ we want to estimate.

\noindent
Once the edge set \textbf{E} is estimated and we have the set of constraints for the elements in the precision matrix, we have to solve the following constrained optimization problem:
\begin{equation}
\begin{split}
&\max_{\PrecisionN} \Big{\{} \text{log}(\text{det}(\PrecisionN)) - \text{trace}(\SampleCov\PrecisionN) \Big{\}} \\
&\text{ s.t.} \\
&\theta_{ij} = \theta_{ji} = 0 \text{ if edge } (i,j) \notin\hat{\text{\textbf{E}}}
\end{split}
\end{equation}
This problem can be rewritten in Lagrangian form as:
\begin{equation}
	\argmax_{\PrecisionN} \Big{\{} \text{log}(\text{det}(\PrecisionN)) - \text{trace}(\SampleCov\PrecisionN) - \sum_{(i,j)\notin\mathbf{E}}\gamma_{ij}\theta_{ij}\Big{\}}
\end{equation}
To solve this optimization problem, we use the algorithm proposed by Hastie et al. \cite{bookESL2009} to maximize the constrained log-likelihood and produce a (sparse) estimate $\hat{\PrecisionN}$, given the network structure already estimated in the previous step.

\noindent
Pseudo-code for the whole procedure is reported in \hyperref[alg3]{\textbf{Algorithm 3}}
\begin{algorithm} \label{alg3}
\caption{\emph{2S-gelnet}}
\label{alg5}
\begin{algorithmic}
\FOR{i=1,..,p}
\STATE Fit an elastic net penalized regression using coordinate descent procedure in order to estimate the vector of coefficients $\Regc_i$
\ENDFOR
\STATE Use each $\hat{\Regc}_i$ estimated coefficients' vector in combination with an AND or OR rule to estimate the edge set \textbf{E}
\STATE Estimate the precision matrix $\PrecisionN$ with the constraints given by the estimated edge set. That is, the elements of the estimated precision are constrained to zero if the corresponding edge is missing
\end{algorithmic}
\end{algorithm}

\newpage
\section{Methodological Set-Up}\label{sec:setup}

\subsection{Simulation Set-Up}
\noindent
Here, we  briefly describe the simulation set-up to test the performances of the three estimators described in section \ref{sec:description}. For the \emph{CR-gelnet} we test both the L2 method and MinEl method. For the \emph{2S-gelnet} approach we test the AND and OR rules. In addition, as a benchmark, we also consider the  graphical lasso \cite{glasso08}.
We test the estimators on 7 network structures/graphs with 30 nodes each (see appendix \ref{app:1} for adjacency matrices).
\begin{enumerate}
	\itemsep-0.5em 
	\item[] Net1: \textbf{Scale-Free}
	\item[] Net2: \textbf{Random}
	\item[] Net3: \textbf{Hub}
	\item[] Net4: \textbf{Cluster}
	\item[] Net5: \textbf{Band}
	\item[] Net6: \textbf{Small-World}
	\item[] Net7: \textbf{Core-Periphery}
\end{enumerate}

\noindent
For each network structure, 30 datasets are randomly generated from a multivariate normal distribution. We consider both 1000 and 200 sample sizes in order to check the performances in low and high-dimensional case, respectively. We rely on the R package \emph{huge}  to generate the sparse precision matrices that represent the given network structures (for topologies 1-5, parameters: $v$=0.3; $u$=0.1). Moreover, we also consider small-world and core-periphery topologies, as they are well-known structures, and to generate simulated data we rely on the algorithms proposed by Watts and Strogatz \cite{smallworld98} and by Torri et al. \cite{Torri18}.  After specifying the precision matrix, we invert it and thus use the covariance matrix as input to generate data from a multivariate Gaussian distribution.
For each dataset,  we estimate the best model using BIC or 5-fold cross-validation   (see for example \cite{glasso08}, \cite{SHDD11}, \cite{Torri18} and \cite{ebic10}) with the following grid of tuning parameters:
\begin{itemize}
	\item $\mathbf{\alpha}$ : 41 equally spaced values between 0 and 1
	\item $\mathbf{\lambda}$ : 101 equally spaced values between 0 and 0.4
\end{itemize}
 We notice that it might happen that multiple equivalent optima are found for different combinations of $(\alpha,\lambda)$. In such situations, we consider the one with the minimum $\alpha$.

\vspace{0.5cm}
\subsection{Performance measures}
\noindent
We consider a correctly identified edge as a true positive, while a false positive is a missing edge that is incorrectly included in the estimated edge set. Thus, a true negative is a missing edge that is correctly excluded from the estimated edge set, while false negative is an existing edge that is not identified. We use Receiver Operator Characteristic (ROC) curves, accuracy and F$_1$-score to evaluate the binary classification performances.

\noindent
ROC curves report the performance of a binary classifier in terms of false positive rate (FPR) and true positive rate (TPR). FPR is the ratio between false positives and all real negatives, while TPR is defined as the ratio of true positives over total real positives. Different estimates are produced for each combination of the parameters $\alpha$ and $\lambda$. Then, each estimate is plotted on a plane accordingly to its FPR and TPR, reported on x-axis and y-axis, respectively. In each ROC curve, a single point is an estimate of the underlying graph for a given value of $\lambda$. For the estimators based on elastic net penalty, we plot only the estimated models, given the optimally selected value of $\alpha$, as plotting all points  for each $\alpha$ would result in lack of clarity. So here the varying parameters is $\lambda$ and not a probability threshold as with common binary classifiers.

\noindent
 Symbols "o" and "x" are used  to plot ROC curves for BIC and cross-validation, respectively. The optimally selected models are then reported in black instead of red.

\noindent
Accuracy and F$_1$-score are defined as follows:
\begin{itemize}
    \item Accuracy = $\frac{\text{True positives}+\text{True Negatives}}{\text{Positives}+\text{Negatives}}$
    \item F$_1$-score = $2\frac{\text{Precision} \cdot \text{Recall}}{\text{Precision} + \text{Recall}}$ where:
    \begin{itemize}
        \item Precision = $\frac{\text{True positives}}{\text{True Positives}+\text{False Positives}}$
        \item Recall = $\frac{\text{True positives}}{\text{True Positives}+\text{False Negatives}}$
    \end{itemize}
\end{itemize}
Both measures are bounded between 0 and 1. The higher the value, the better the classification of the estimator. While accuracy is generally easier to interpret,  F$_1$-score is more suitable when there is imbalance among classes, that is when the ratio of the number of actual edges over the total number of edges in the hypothetical complete graph is far from 0.5.
\noindent
Finally, we also evaluate the Frobenius distance among estimated and true partial correlation matrices as a measure of numerical accuracy of the estimates. One can retrieve a partial correlation matrix \textbf{P} or $\hat{\text{\textbf{P}}}$ from the true and estimated precision matrix, respectively, using $p_{ij} = \frac{-\theta_{ij}}{\sqrt{\theta_{ii}\theta_{jj}}}$ for $i,j = 1,2,..,k$. We follow a common convention and set $p_{ii} = 0$ for $i = 1,2,..,k$.
The Frobenius distance is then defined as:
\begin{equation}
    ||\text{\textbf{P}}-\hat{\text{\textbf{P}}}||_\text{F} = \sqrt{\sum_{i=1}^p\sum_{j=1}^p |p_{ij}-\hat{p}_{ij}|^2}
\end{equation}
The closer the Frobenius distance is to 0, the better the estimate.
\newpage
\section{Simulation Analysis} \label{sec:results}
\noindent
In this section, we analyse the performance of the three elastic net algorithms and the graphical lasso (\emph{glasso}) on the seven network configurations (see Section \ref{sec:setup} and appendix \ref{app:1}), using both BIC and 5-fold cross validation. Figure \ref{fig:roc} displays the ROC curves for the simulations with sample size $n=$1000, while Tables \ref{acc1000bic}, \ref{acc1000cv} and \ref{f1m1000bic}, \ref{f1m1000cv} report the mean values and standard deviations of accuracy and F$_1$-score for the BIC and 5-fold cross validation, respectively. Tables \ref{fd1000bic} and \ref{fd1000cv} report the mean values of Frobenius distance in the low-dimensional case. Tables \ref{alp1000bic}, \ref{alp1000cv}, \ref{alp200bic} and \ref{alp200cv} in the appendix \ref{app:21} and \ref{app:22} display the average value of the optimally selected $\alpha$.
Tables \ref{acc200bic}, \ref{acc200cv}, \ref{f1m200bic}, \ref{f1m200cv}, \ref{fd200bic} and \ref{fd200cv} report the mean values of accuracy, F$_1$-score and Frobenius distance in the high-dimensional case ($n$=200). Figure \ref{fig:roc200} then displays the ROC curves in the high-dimensional case.

\begin{figure}[h!]
\begin{subfigure}{1\textwidth}
    \hspace{-1cm}
  \includegraphics[width=1.2\linewidth]{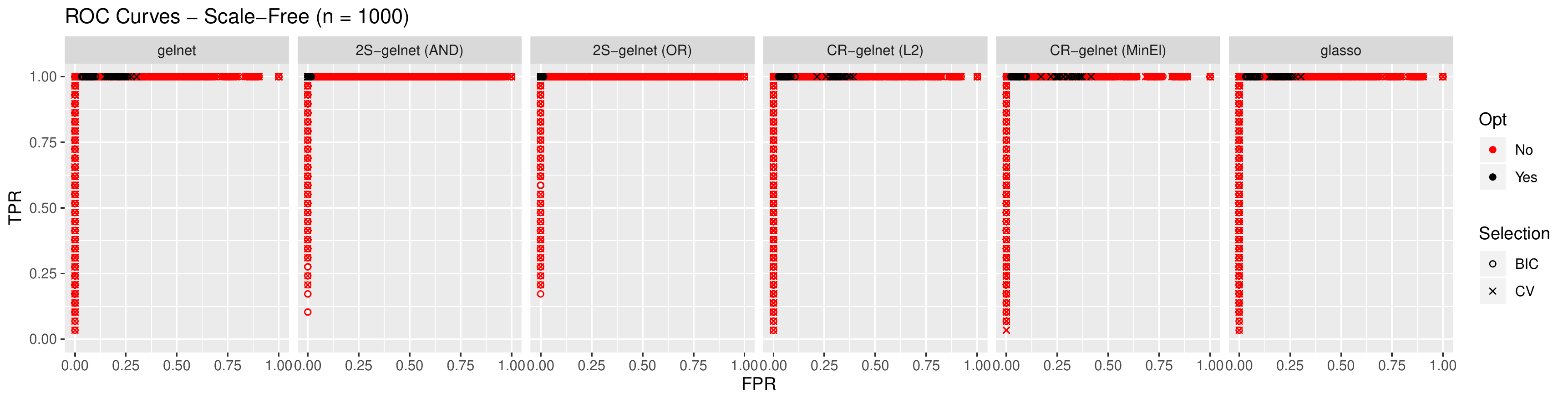}
\end{subfigure}%
\newline
\begin{subfigure}{1\textwidth}
\hspace{-1cm}
  \includegraphics[width=1.2\linewidth]{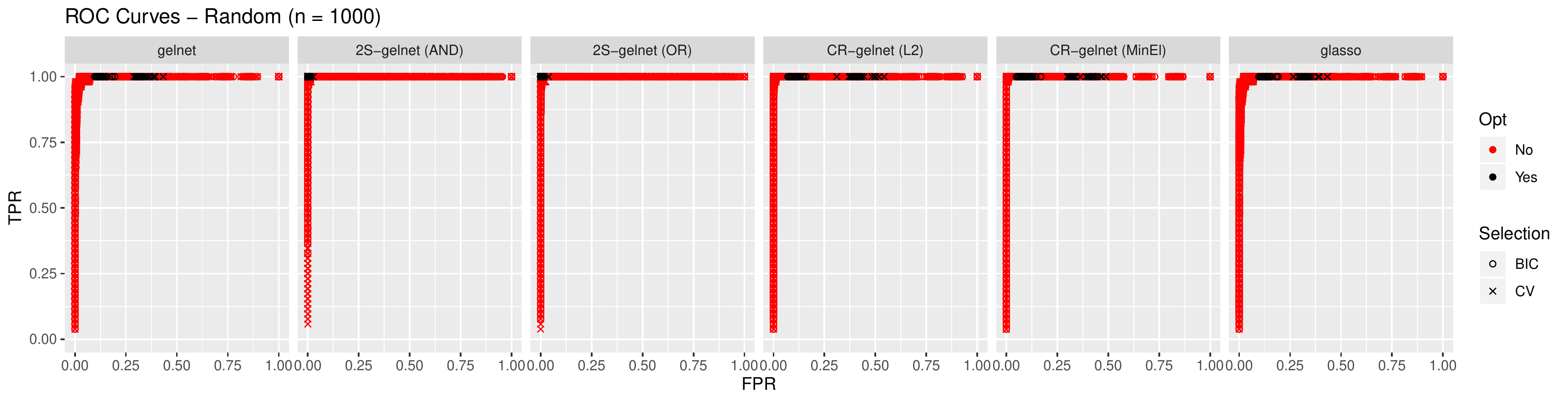}
\end{subfigure}
\newline
\begin{subfigure}{1\textwidth}
\hspace{-1cm}
  \includegraphics[width=1.2\linewidth]{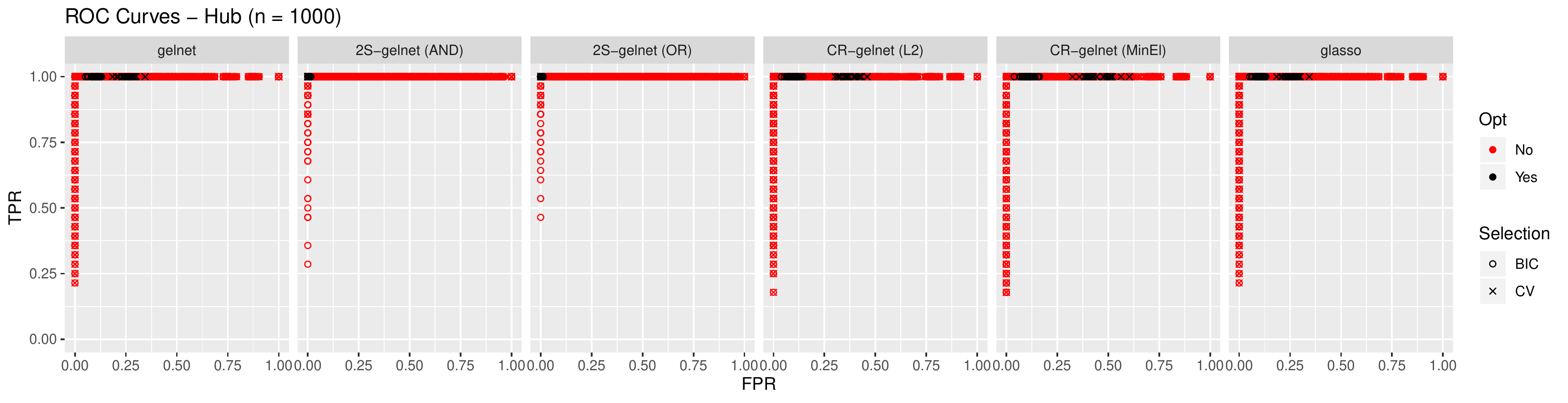}
\end{subfigure}
\end{figure}
\clearpage
\begin{figure}[ht]\ContinuedFloat
\begin{subfigure}{1\textwidth}
\hspace{-1cm}
  \includegraphics[width=1.2\linewidth]{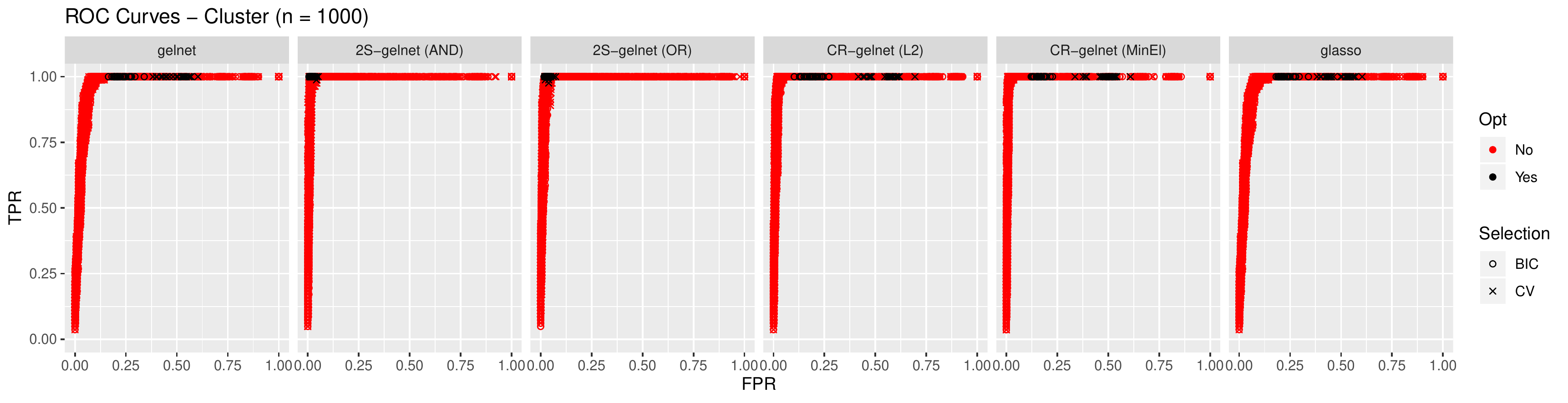}
\end{subfigure}
\begin{subfigure}{1\textwidth}
\hspace{-1cm}
  \includegraphics[width=1.2\linewidth]{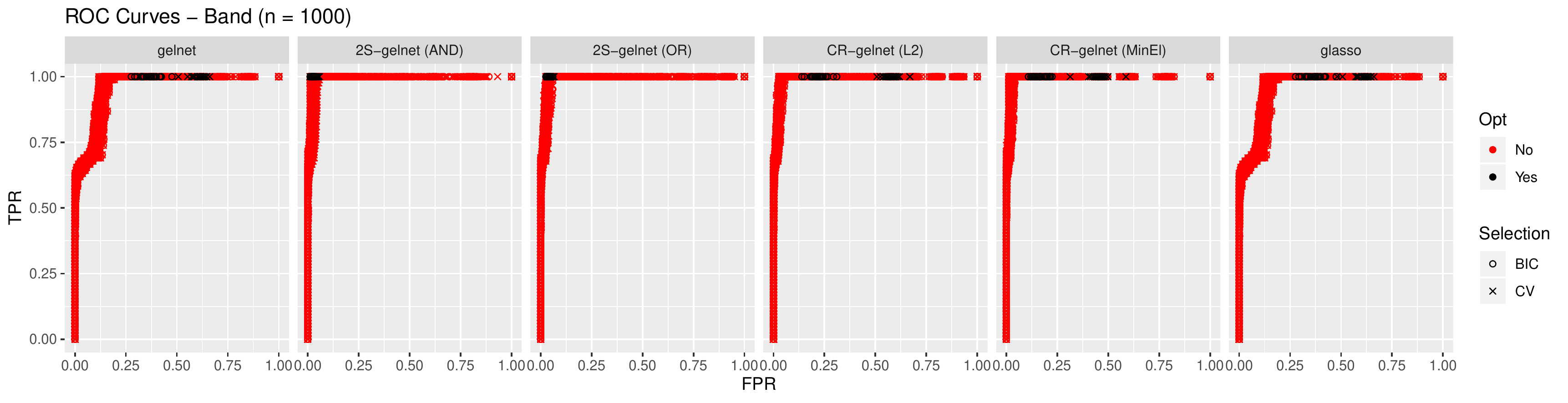}
\end{subfigure}
\newline
\begin{subfigure}{1\textwidth}
\hspace{-1cm}
  \includegraphics[width=1.2\linewidth]{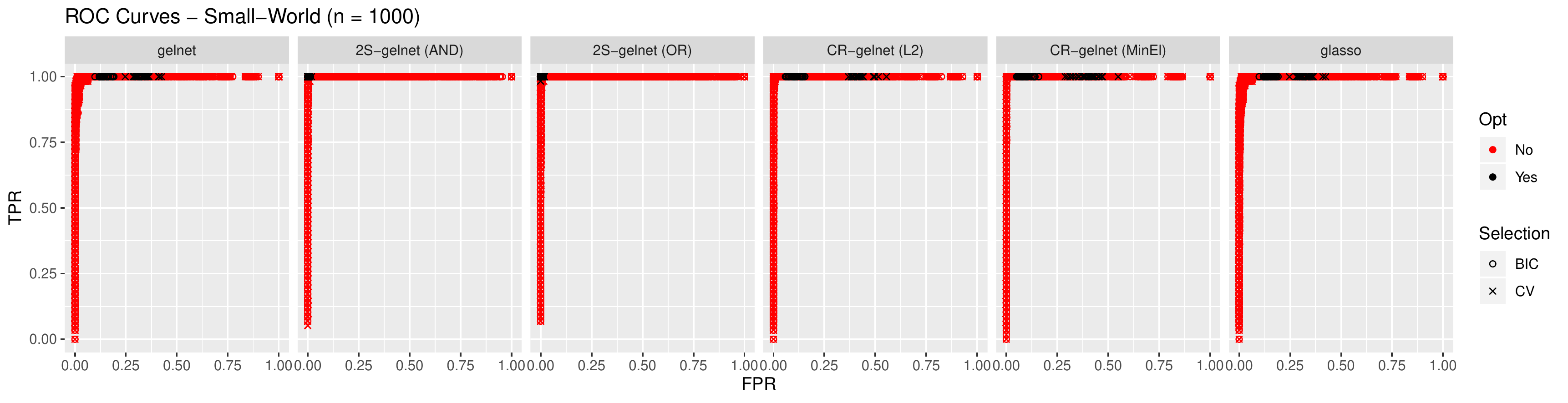}
\end{subfigure}
\newline
\begin{subfigure}{1\textwidth}
\hspace{-1cm}
  \includegraphics[width=1.2\linewidth]{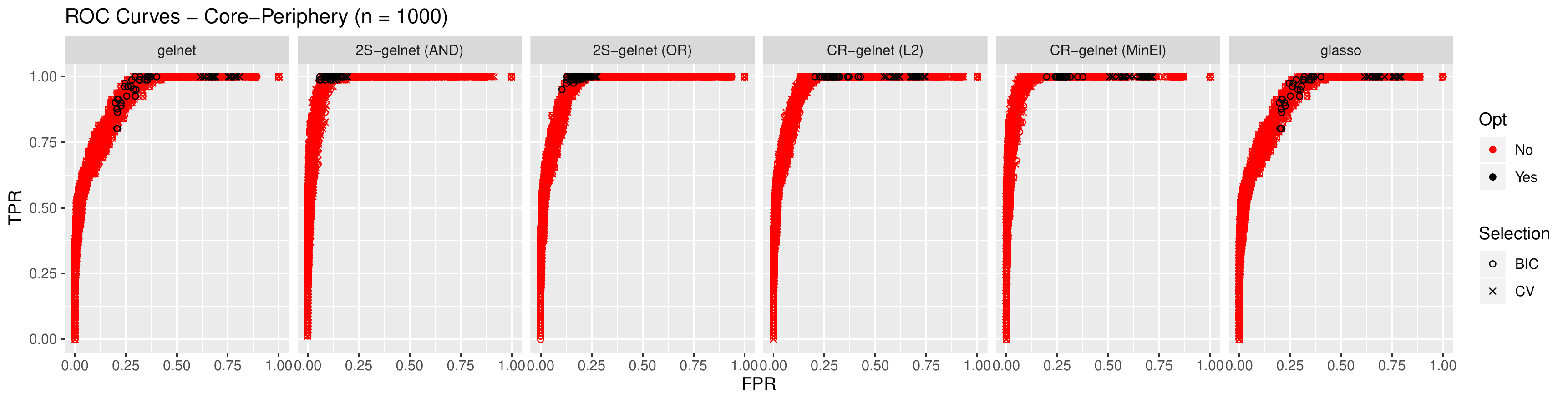}
\end{subfigure}
\caption{ROC curves for seven network structures (in rows) for gelnet (column 1), 2S-gelnet (columns 2 \& 3), CR-gelnet (columns 4 \& 5) and glasso (column 6) - (n=1000)}
\label{fig:roc}
\end{figure}
\clearpage

\noindent
From the ROC curves in Figure \ref{fig:roc}, we see that the \emph{2S-gelnet} in columns 2 and 3 performs always better than the other methods. Both using BIC criterion and cross-validation, the optimally selected models are the closest to the top-left corner of $[0,1]\times[0,1]$ square (high TPR, low FPR). Also notice that the optimal estimates of the 2-stages approach \emph{2S-gelnet}  are very close both using BIC and cross-validation. When considering the other methods (i.e. \emph{gelnet} - column 1, \emph{CR-gelnet} - columns 4 and 5 and \emph{glasso} - column 6) the BIC criterion leads to better models than the cross-validation in terms of classification performance, as the optimal models (i.e. black circle) are closer to the top-left corner of the ROC curve than the models selected by cross-validation (i.e. black cross).

\vspace{2cm}
\begin{table}[!h]
\centering
\begin{tabular}{rrrrrrr}
  \hline
 & gelnet & 2S-gelnet & 2S-gelnet & CR-gelnet & CR-gelnet & glasso \\
 &  & (AND) & (OR) & (L2) & (MinEl) &  \\
  \hline
Scale-Free       & 0.9394 & 0.9955 & \textbf{0.9962} & 0.9495 & 0.9456 & 0.9399 \\ 
                 & (0.0179) & (0.0048) & (0.0039) & (0.0175) & (0.0206) & (0.0175) \\ 
Random           & 0.8818 & 0.9958 & \textbf{0.9960} & 0.9084 & 0.9186 & 0.8818 \\ 
                 & (0.0207) & (0.0039) & (0.0036) & (0.0208) & (0.0262) & (0.0200) \\ 
Hub              & 0.9072 & \textbf{0.9972} & 0.9962 & 0.9088 & 0.9004 & 0.9072 \\ 
                 & (0.0185) & (0.0033) & (0.0041) & (0.0235) & (0.0267) & (0.0185) \\ 
Cluster          & 0.8153 & \textbf{0.9847} & 0.9742 & 0.8474 & 0.8665 & 0.8140 \\ 
                 & (0.0309) & (0.0065) & (0.0078) & (0.0359) & (0.0212) & (0.0293) \\ 
Band             & 0.7099 & \textbf{0.9795} & 0.9700 & 0.8241 & 0.8657 & 0.7100 \\ 
                 & (0.0362) & (0.0051) & (0.0056) & (0.0296) & (0.0271) & (0.0362) \\ 
Small-World      & 0.8740 & \textbf{0.9963} & 0.9959 & 0.9057 & 0.9150 & 0.8748 \\ 
                 & (0.0201) & (0.0031) & (0.0035) & (0.0241) & (0.0256) & (0.0202) \\ 
Core-Periphery   & 0.7606 & \textbf{0.9155} & 0.8629 & 0.7588 & 0.7710 & 0.7582 \\ 
                 & (0.0409) & (0.0185) & (0.0214) & (0.0416) & (0.0266) & (0.0416) \\ 
   \hline
\end{tabular}
\caption{Average Accuracy (Std. Dev. in brackets) - BIC calibration ($n$ = 1000)}
\label{acc1000bic}
\end{table}
\clearpage
\begin{table}[ht]
\centering
\begin{tabular}{rrrrrrr}
  \hline
 & gelnet & 2S-gelnet & 2S-gelnet & CR-gelnet & CR-gelnet & glasso \\
 &  & (AND) & (OR) & (L2) & (MinEl) &  \\
  \hline
Scale-Free       & 0.8013 & \textbf{0.9994} & 0.9991 & 0.7118 & 0.7239 & 0.8034 \\ 
                 & (0.0373) & (0.0013) & (0.0021) & (0.0340) & (0.0524) & (0.0360) \\ 
Random           & 0.7038 & \textbf{0.9968} & 0.9949 & 0.6270 & 0.6655 & 0.7145 \\ 
                 & (0.0338) & (0.0049) & (0.0066) & (0.0433) & (0.0506) & (0.0339) \\ 
Hub              & 0.7545 & \textbf{0.9996} & \textbf{0.9996} & 0.6517 & 0.5707 & 0.7595 \\ 
                 & (0.0338) & (0.0011) & (0.0012) & (0.0422) & (0.0598) & (0.0313) \\ 
Cluster          & 0.5946 & \textbf{0.9803} & 0.9693 & 0.5523 & 0.6017 & 0.6038 \\ 
                 & (0.0475) & (0.0093) & (0.0104) & (0.0520) & (0.0444) & (0.0505) \\ 
Band             & 0.5094 & \textbf{0.9734} & 0.9659 & 0.5315 & 0.6347 & 0.5113 \\ 
                 & (0.0251) & (0.0092) & (0.0072) & (0.0282) & (0.0335) & (0.0302) \\ 
Small-World      & 0.7133 & \textbf{0.9980} & 0.9962 & 0.6341 & 0.6628 & 0.7182 \\ 
                 & (0.0336) & (0.0035) & (0.0048) & (0.0421) & (0.0533) & (0.0319) \\ 
Core-Periphery   & 0.4226 & \textbf{0.8861} & 0.8258 & 0.4654 & 0.4772 & 0.4305 \\ 
                 & (0.0473) & (0.0224) & (0.0284) & (0.0499) & (0.0520) & (0.0441) \\ 

\hline
\end{tabular}
\caption{Average Accuracy (Std. Dev. in brackets) - 5-CV calibration ($n$ = 1000)}
\label{acc1000cv}
\end{table}

\noindent
Tables \ref{acc1000bic} and \ref{acc1000cv} report the average accuracy (standard deviation in brackets) for the BIC and 5-fold cross validation, respectively. Bold values indicate best result. Notice that the 2 stages procedure \emph{2S-gelnet} consistently outperforms all other methods in terms of accuracy. When we use BIC calibration, \emph{2S-gelnet} (OR) is better than AND rule for scale-free and random networks, while \emph{2S-gelnet} (AND) is always better or equal to \emph{2S-gelnet} (OR) when 5-fold cross-validation is considered. \emph{CR-gelnet} tends to outperform \emph{glasso} and \emph{gelnet} for BIC calibration (see Table \ref{acc1000bic}), while reporting mixed results for 5-fold cross-validation (see Table \ref{acc1000cv}).

\newpage
\begin{table}[ht]
\centering
\begin{tabular}{rrrrrrr}
  \hline
 & gelnet & 2S-gelnet & 2S-gelnet & CR-gelnet & CR-gelnet & glasso \\
 &  & (AND) & (OR) & (L2) & (MinEl) &  \\ 
  \hline
Scale-Free       & 0.6932 & 0.9683 & \textbf{0.9734} & 0.7315 & 0.7186 & 0.6949 \\ 
                 & (0.0647) & (0.0328) & (0.0270) & (0.0675) & (0.0794) & (0.0635) \\ 
Random           & 0.6712 & 0.9829 & \textbf{0.9838} & 0.7256 & 0.7507 & 0.6711 \\ 
                 & (0.0374) & (0.0155) & (0.0145) & (0.0445) & (0.0590) & (0.0362) \\ 
Hub              & 0.5853 & \textbf{0.9790} & 0.9720 & 0.5922 & 0.5715 & 0.5853 \\ 
                 & (0.0519) & (0.0238) & (0.0294) & (0.0669) & (0.0693) & (0.0519) \\ 
Cluster          & 0.6731 & \textbf{0.9612} & 0.9362 & 0.7151 & 0.7397 & 0.6713 \\ 
                 & (0.0356) & (0.0157) & (0.0180) & (0.0486) & (0.0304) & (0.0332) \\ 
Band             & 0.5726 & \textbf{0.9497} & 0.9280 & 0.6889 & 0.7440 & 0.5727 \\ 
                 & (0.0303) & (0.0118) & (0.0123) & (0.0360) & (0.0387) & (0.0303) \\ 
Small-World      & 0.6809 & \textbf{0.9865} & 0.9852 & 0.7419 & 0.7622 & 0.6822 \\ 
                 & (0.0345) & (0.0111) & (0.0126) & (0.0499) & (0.0547) & (0.0348) \\ 
Core-Periphery   & 0.5982 & \textbf{0.8160} & 0.7311 & 0.6095 & 0.6203 & 0.5959 \\ 
                 & (0.0318) & (0.0330) & (0.0298) & (0.0396) & (0.0272) & (0.0318) \\ 
\hline
\end{tabular}
\caption{Average F$_1$-score (Std. Dev. in brackets) - BIC calibration ($n$ = 1000)}
\label{f1m1000bic}
\end{table}
\vspace{1cm}
\begin{table}[h!]
\centering
\begin{tabular}{rrrrrrr}
  \hline
 & gelnet & 2S-gelnet & 2S-gelnet & CR-gelnet & CR-gelnet & glasso \\
 &  & (AND) & (OR) & (L2) & (MinEl) &  \\ 
  \hline
Scale-Free       & 0.4065 & \textbf{0.9955} & 0.9934 & 0.3183 & 0.3311 & 0.4088 \\ 
                 & (0.0460) & (0.0098) & (0.0146) & (0.0261) & (0.0442) & (0.0445) \\ 
Random           & 0.4483 & \textbf{0.9871} & 0.9796 & 0.3925 & 0.4200 & 0.4575 \\ 
                 & (0.0271) & (0.0190) & (0.0246) & (0.0271) & (0.0370) & (0.0283) \\ 
Hub              & 0.3467 & \textbf{0.9971} & \textbf{0.9971} & 0.2719 & 0.2333 & 0.3510 \\ 
                 & (0.0314) & (0.0080) & (0.0092) & (0.0237) & (0.0256) & (0.0298) \\ 
Cluster          & 0.4836 & \textbf{0.9508} & 0.9253 & 0.4589 & 0.4879 & 0.4896 \\ 
                 & (0.0305) & (0.0220) & (0.0231) & (0.0300) & (0.0301) & (0.0322) \\ 
Band             & 0.4408 & \textbf{0.9360} & 0.9191 & 0.4523 & 0.5149 & 0.4420 \\ 
                 & (0.0131) & (0.0208) & (0.0155) & (0.0147) & (0.0236) & (0.0163) \\ 
Small-World      & 0.4835 & \textbf{0.9927} & 0.9861 & 0.4233 & 0.4449 & 0.4878 \\ 
                 & (0.0286) & (0.0126) & (0.0172) & (0.0263) & (0.0387) & (0.0276) \\ 
Core-Periphery   & 0.3930 & \textbf{0.7671} & 0.6830 & 0.4118 & 0.4175 & 0.3962 \\ 
                 & (0.0194) & (0.0355) & (0.0362) & (0.0230) & (0.0255 )& (0.0181) \\ 
   \hline
\end{tabular}
\caption{Average F$_1$-score (Std. Dev. in brackets) - 5-CV calibration ($n$ = 1000)}
\label{f1m1000cv}
\end{table}
\clearpage

\noindent
When considering the mean F$_1$-scores, reported in Tables \ref{f1m1000bic} and \ref{f1m1000cv} for BIC calibration and 5-fold cross-validation, respectively,  results are qualitatively similar, with \emph{2S-gelnet} clearly outperforming all other methods, reaching values very close to 1. The other methods instead report quite unsatisfactory performances in comparison, especially \emph{CR-gelnet} for 5-fold cross validation.

\noindent
Finally, Tables \ref{fd1000bic} and \ref{fd1000cv} report  the average Frobenius distance between the theoretical partial correlation matrix and the estimated one. The lower the value, the closer is the estimate to the true model. Similarly to the classification performances, the values in these tables show that the \emph{2S-gelnet} performs better then other methods, using both BIC criterion approach and 5-fold cross-validation. 
\vspace{1cm}
~
\begin{table}[!h]
\centering
\begin{tabular}{lrrrrrr}
  \hline
 & gelnet & 2S-gelnet & 2S-gelnet & CR-gelnet & CR-gelnet & glasso \\
 &  & (AND) & (OR) & (L2) & (MinEl) &  \\
  \hline
Scale-Free       & 0.4169 & 0.2369 & \textbf{0.2276} & 0.4372 & 0.4352 & 0.4174 \\ 
                 & (0.0472) & (0.0628) & (0.0582) & (0.0419) & (0.0483) & (0.0469) \\ 
Random           & 0.5965 & 0.2758 & \textbf{0.2712} & 0.5764 & 0.5532 & 0.5958 \\ 
                 & (0.0418) & (0.0397) & (0.0368) & (0.0462) & (0.0464) & (0.0427) \\ 
Hub              & 0.3084 & \textbf{0.1884} & 0.1999 & 0.3446 & 0.3870 & 0.3084 \\ 
                 & (0.0312) & (0.0572) & (0.0651) & (0.0261) & (0.0339) & (0.0312) \\ 
Cluster          & 0.9486 & \textbf{0.3794} & 0.3987 & 0.8028 & 0.7262 & 0.9454 \\ 
                 & (0.0740) & (0.0431) & (0.0395) & (0.0654) & (0.0550) & (0.0740) \\ 
Band             & 1.1918 & \textbf{0.3612} & 0.3727 & 0.9540 & 0.8147 & 1.1919 \\ 
                 & (0.1045) & (0.0405) & (0.0432) & (0.0879) & (0.0752) & (0.1045) \\ 
Small-World      & 0.5693 & 0.2815 & \textbf{0.2780} & 0.5681 & 0.5426 & 0.5699 \\ 
                 & (0.0436) & (0.0347) & (0.0334) & (0.0467) & (0.0480) & (0.0409) \\ 
Core-Periphery   & 2.0539 & \textbf{0.5885} & 0.7003 & 1.4341 & 1.1791 & 2.0477 \\ 
                 & (0.1492) & (0.1001) & (0.1718) & (0.1394) & (0.0943) & (0.1510) \\ 
   \hline
\end{tabular}
\caption{Average Frobenius distance (Std. Dev. in brackets) - BIC calibration ($n$ = 1000)}
\label{fd1000bic}
\end{table}

\clearpage

\begin{table}[ht]
\centering
\begin{tabular}{rrrrrrr}
  \hline
 & gelnet & 2S 2S-gelnet & 2S-gelnet & CR-gelnett & CR-gelnett & glasso \\
 &  & (AND) & (OR) & (L2) & (MinEl) &  \\
  \hline
Scale-Free       & 0.3937 & \textbf{0.1828} & 0.1870 & 0.4148 & 0.4419 & 0.3928 \\ 
                 & (0.0397) & (0.0405) & (0.0466) & (0.0395) & (0.0473) & (0.0391) \\ 
Random           & 0.5160 & \textbf{0.2608} & 0.2708 & 0.5020 & 0.5224 & 0.5154 \\ 
                 & (0.0288) & (0.0432) & (0.0455) & (0.0321) & (0.0334) & (0.0291) \\ 
Hub              & 0.4087 & 0.1374 & \textbf{0.1364} & 0.4427 & 0.5340 & 0.4062 \\ 
                 & (0.0308) & (0.0382) & (0.0393) & (0.0343) & (0.0451) & (0.0325) \\ 
Cluster          & 0.7179 & \textbf{0.3944} & 0.4260 & 0.6405 & 0.6433 & 0.7219 \\ 
                 & (0.0360) & (0.0588) & (0.0808) & (0.0371) & (0.0364) & (0.0354) \\ 
Band             & 0.8113 & \textbf{0.3688} & 0.3792 & 0.6550 & 0.6282 & 0.8143 \\ 
                 & (0.0493) & (0.0416) & (0.0443) & (0.0478) & (0.0479) & (0.0509) \\ 
Small-World      & 0.5045 & \textbf{0.2606} & 0.2757 & 0.5010 & 0.5254 & 0.5042 \\ 
                 & (0.0362) & (0.0293) & (0.0463) & (0.0405) & (0.0468) & (0.0360) \\ 
Core-Periphery   & 1.1909 & \textbf{0.6252} & 0.6640 & 0.8655 & 0.8096 & 1.2073 \\ 
                 & (0.1027) & (0.0978) & (0.0857) & (0.0652) & (0.0596) & (0.0994) \\ 
   \hline
\end{tabular}
\caption{Average Frobenius distance (Std. Dev. in brackets) - 5-CV calibration ($n$ = 1000)}
\label{fd1000cv}
\end{table}
\vspace{1cm}

\noindent
Appendix \ref{app:22} reports the results with sample size of 200. We see that the results in the high-dimensional case are similar to the ones discussed in this section when considering the classification performances. Looking at accuracy and F$_1$-scores in Tables \ref{acc200bic}, \ref{acc200cv} and Tables \ref{f1m200bic}, \ref{f1m200cv}, they qualitatively hold (with only one exception when \emph{CR-gelnet} (L2) slightly outperforms \emph{2S-gelnet} (AND) in term of F$_1$-score using BIC calibration, see Table \ref{f1m200bic}). Still, if we look at the optimally selected models reported in ROC curves, we don't always have an estimator whose optimal estimates are clearly the highest and leftmost in the $[0,1]\times[0,1]$ square. One difference is that in the high-dimensional situation \emph{2S-gelnet} (AND) is better \emph{2S-gelnet} (OR) only when considering cluster, band and small-world networks looking both at average accuracy and F$_1$-score. Looking at the Frobenius distance in Tables \ref{fd200bic} and \ref{fd200cv}, most of the times there is at least one version (between AND/OR rules) of \emph{2S-gelnet} that has better performance than other estimators. There are 2 only two exceptions, where \emph{CR-gelnet} outperforms \emph{2S-gelnet} (see Table \ref{fd200cv}). There are also 4 situations in which only one of version of \emph{2S-gelnet} is outperformed by other algorithms, see Tables \ref{fd200bic} and \ref{fd200cv}.

\noindent
Summing up, empirical results  suggest that the 2-stages estimator \emph{2S-gelnet} tends to have a remarkable performance no matter what the network structure is, the parameter selection procedure used (i.e. BIC or 5-fold cross-validation) and the high/low-dimensional set-up. Empirical results show only 2 cases where both versions of \emph{2S-gelnet} are outperformed by \emph{CR-gelnet}. \emph{CR-gelnet} tends to outperform the \emph{glasso} and \emph{gelnet} when relying on BIC calibration, while it has often unsatisfactory performance for 5-fold cross-validation. \emph{Gelnet} and \emph{glasso} performances are quite close to each other. This is not surprising as the optimal choice for $\alpha$ is often equal to 1, which then results in avoiding the penalty based on the 2-norm and just relying on the 1-norm (see Tables \ref{alp1000bic} and \ref{alp1000cv} in appendix \ref{app:21} for more details). This pattern in the optimal selection of $\alpha$ is replicated also in the high-dimensional case (see Tables \ref{alp200bic} and \ref{alp200cv} in appendix \ref{app:22}), but with a notable exception with the core-periphery structure (i.e. Net 7). In the low-dimensional case, \emph{2S-gelnet} (AND) performs the best in terms of F$_1$-score and it is quite closely followed by \emph{2S-gelnet} (OR), while if we look at the average accuracy it depends on the network structure. In the high-dimensional case there isn't a clear winner between the two versions of \emph{2S-gelnet}, but they still outperform all other methods.

\clearpage

\newpage
\section{The Unites States network of economic sectors} \label{sec:empana}
\noindent
In this section, we focus on the use of  the \emph{2S-gelnet} to estimate the network across different US economic sectors. Such a network could play an important role in the dynamics of aggregate macroeconomic fluctuations  (see for example \cite{ace12}). The network is estimated by using market data log-returns of S\&P 500 sectors' indices, being then the so-called perceived network, as pointed out for example by \cite{anuf15}. This estimated network could differ from the underlying real-world one, even if we had a perfect estimator. Nonetheless the perceived network could be estimated from more readily available market data and it can provide useful information about market expectations relative to economic inter-linkages among sectors.

\begin{figure}[H]
 \hspace*{-2cm}
 \includegraphics[width=0.6\textwidth]{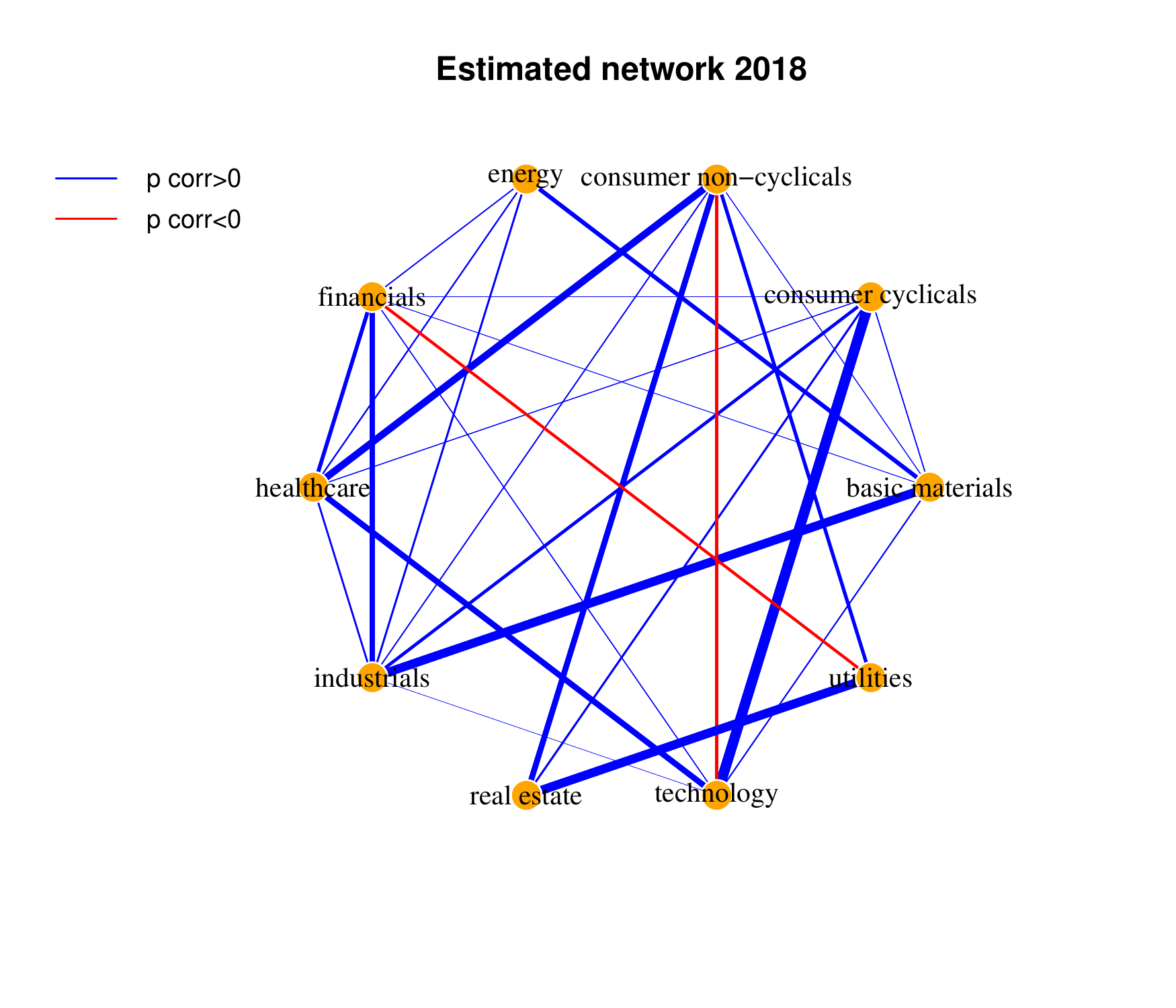}
 \includegraphics[width=0.6\textwidth]{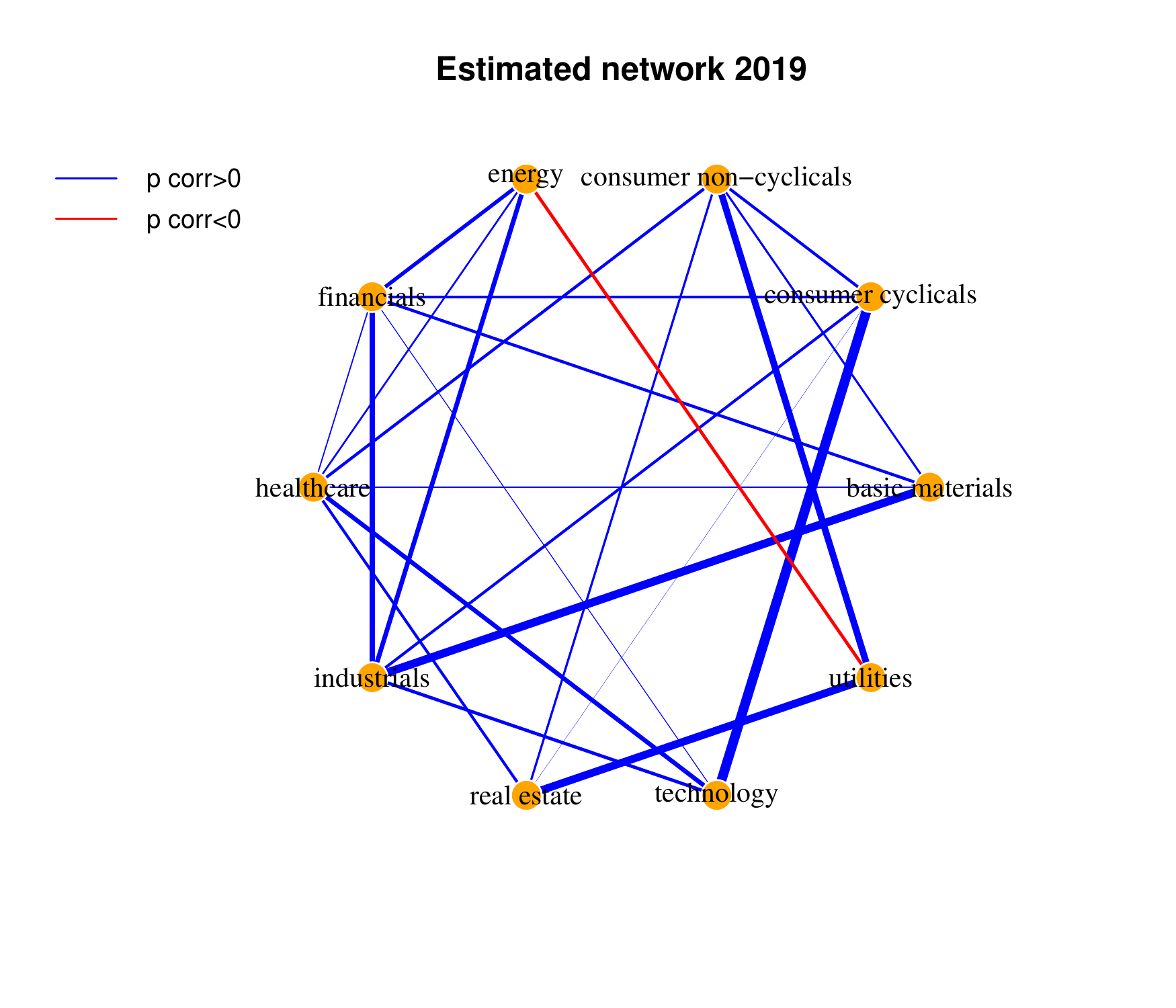}
 ~
 \newline
  \hspace*{-2cm}
 \includegraphics[width=0.6\textwidth]{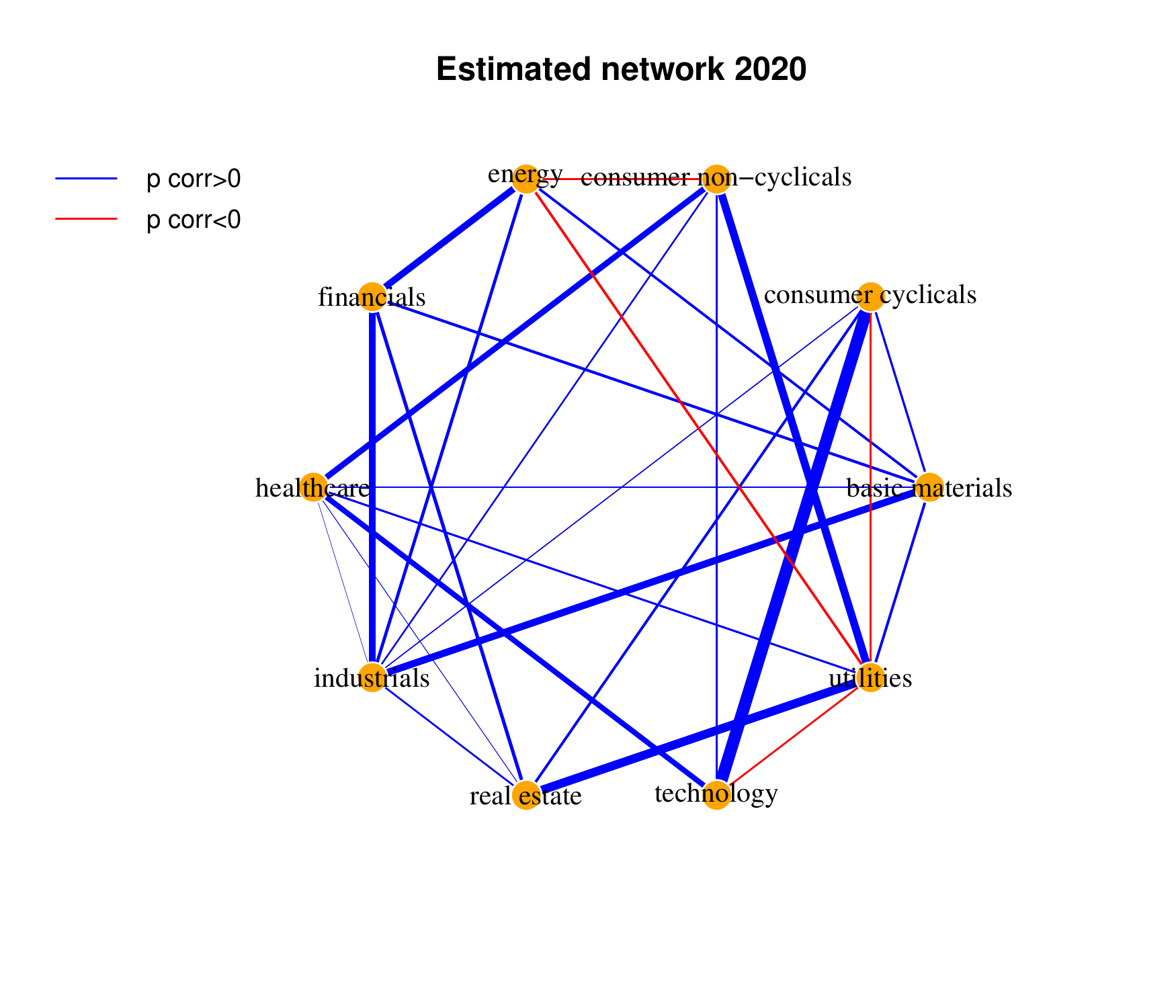}
 \includegraphics[width=0.6\textwidth]{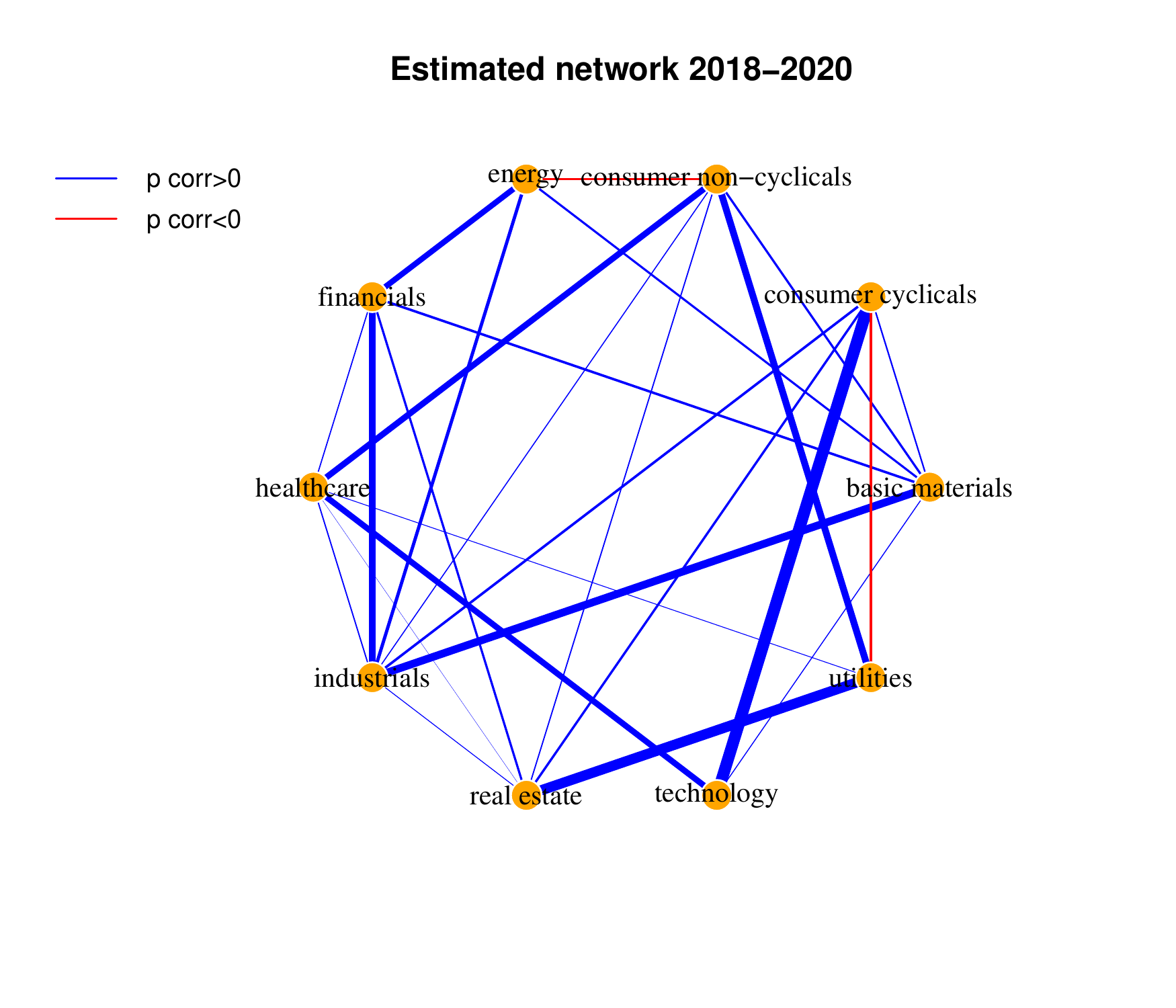}
 \vspace{-2cm}
\caption{Estimated networks using 2S-gelnet (AND)}
\label{fig:est}
\end{figure}

\noindent
We use daily time series of log-returns of the 10 economic sector indices constituents of S\&P 500 in the period 01/01/2018-25/11/2020. To rule out possible serial correlation and  volatility clustering, we don't estimate  the network directly on the daily log-returns. First, we fit auto-regressive models (AR(1)) with generalized auto-regressive conditional heteroskedasticity  (i.e. GARCH(1,1)). Then, we use the residuals as inputs for the graphical models to estimate the perceived network using \emph{2S-gelnet} (AND) with the BIC criterion. We consider both intervals for each year as well the entire  3 years period. Figure \ref{fig:est} displays  the 4 optimal estimated networks. In red, we point out edges corresponding to negative partial correlations, while blue edges represent positive partial correlations. The width of the edge is proportional to the magnitude of the partial correlation estimates.

\newpage
\clearpage

\begin{figure}[ht]
 \hspace*{-2cm}
 \includegraphics[width=0.6\textwidth]{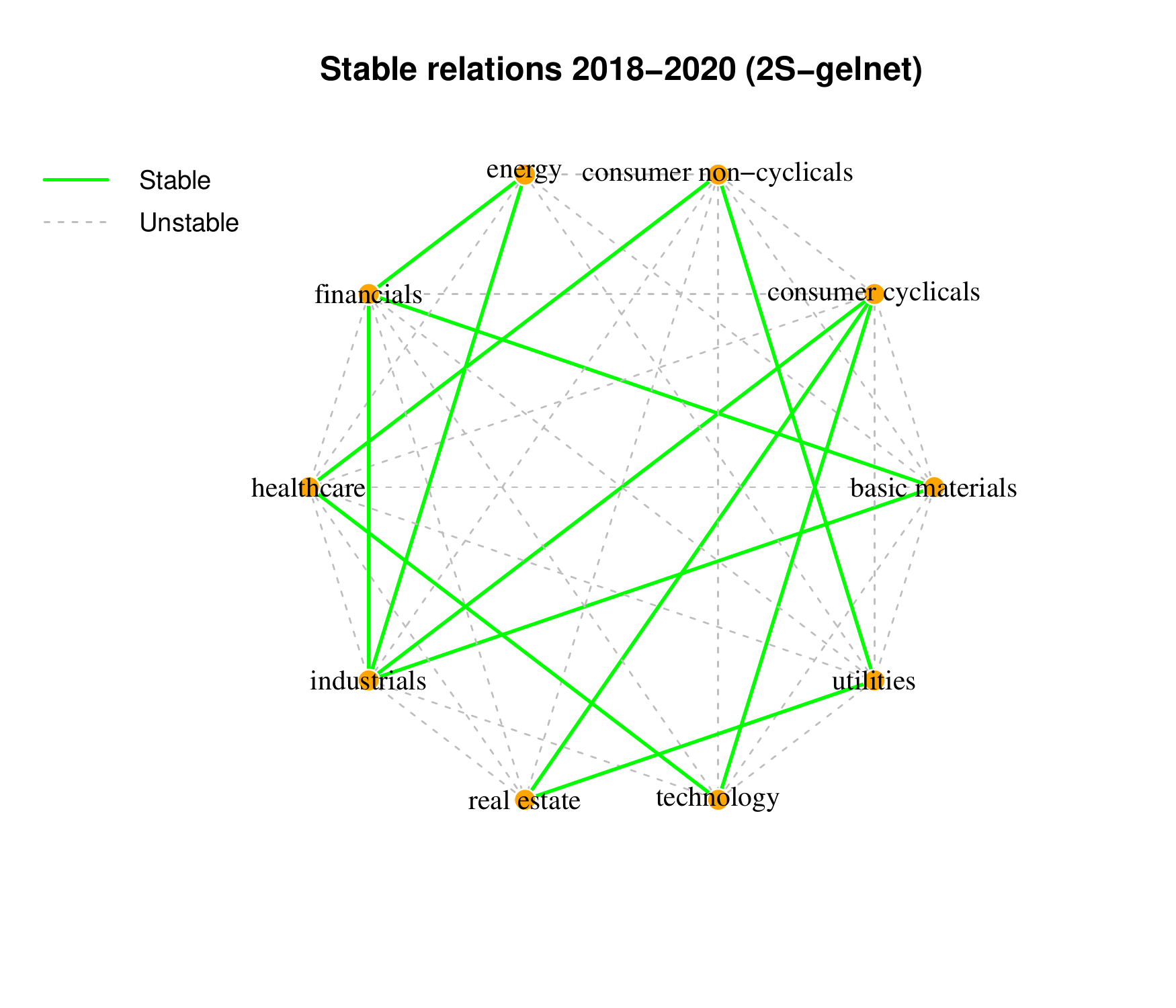}
 \includegraphics[width=0.6\textwidth]{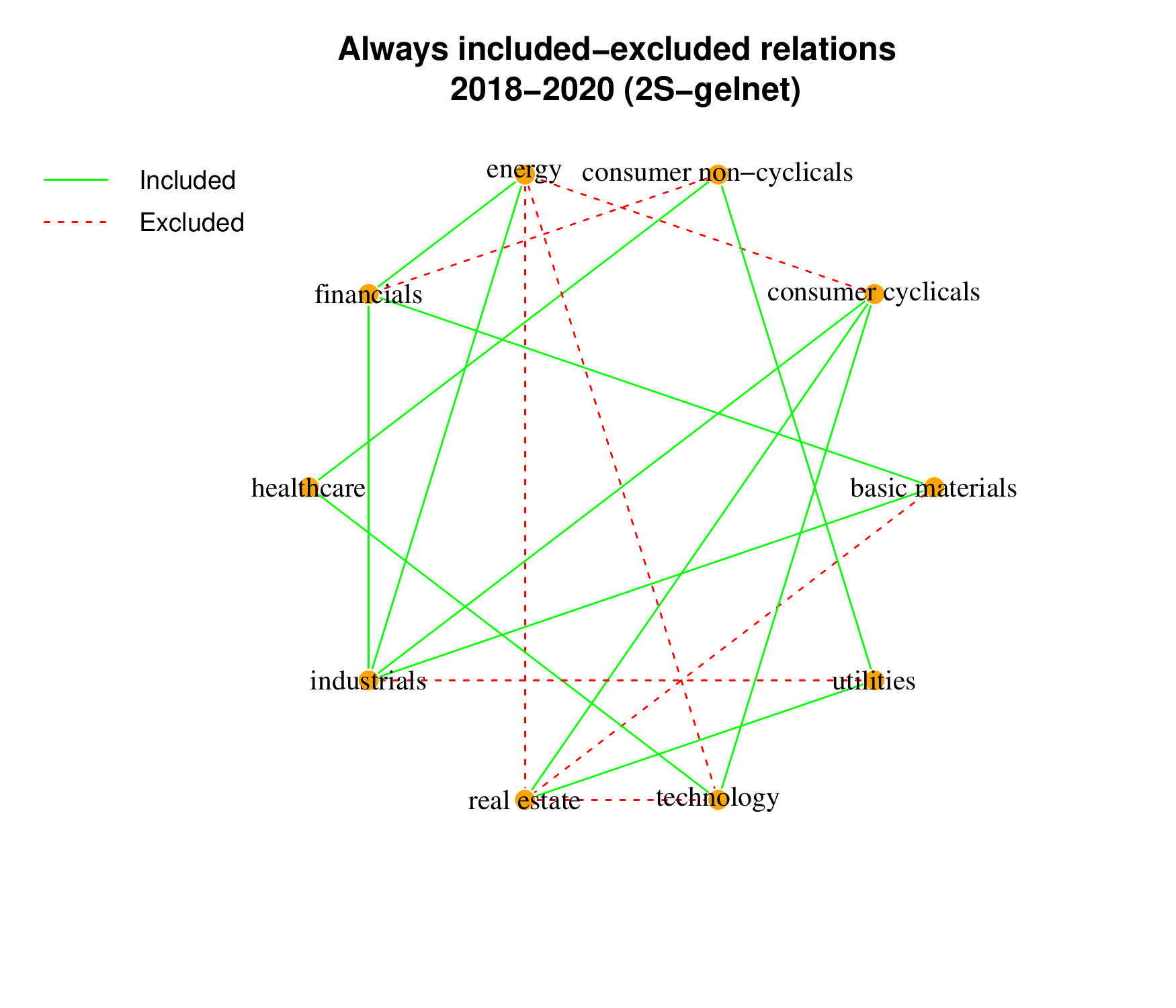}
 \vspace{-2cm}
 \caption{Stable relations (left) and  Always included-excluded relations (right)}
\label{fig:stabinex}
\end{figure}
 ~
 \newline
 Figure \ref{fig:stabinex} reports on the left  the stable relations (in green) among all the dependencies estimated in each of the 3 years and on the right  the always included, or stable, relations (in green) and the always excluded relations (in red). So, for example, the connection between the industrial and basic materials sectors is a stable one across 2018, 2019 and 2020, while the connection between technology and basic materials is not stable because it is only detected in 2018. The estimated networks can then help to better point out hidden relationships to better capture potential spillover effects among sectors. Furthermore, we notice that some connections have been never detected, such as  the one between real estate and technology.
The numbers of estimated links in the entire period has been equal to 26, which is consistent with the 27 links estimated during 2018 and 2020, while only in 2019, we had a lower number of estimated links (i.e. 23). Therefore, the estimated structure suggests that, at least in the perceived network, only a bit more than half of the possible connections are relevant, as we know that the complete graph would have 45 edges in total. 
The increase in the number of connections detected between 2019 and 2020, could also be  possibly attributed  to the Covid-19 pandemic. It is well-known that correlations in financial markets tend to increases during crisis and therefore new connections among sectors could emerge, suggesting other channels of contagions across sectors, which financial actors might want to monitor to cope with risk aversion and a more uncertain scenario. Some new links emerge in the network estimated using 2020 data. In fact, we detect, for example, positive correlations between industrials and real estate and between financial and real estate. 
 The first relation might be associated to a general slowdown of construction industry, at least in the first months since the spread of the pandemic. The second connection may remind us about a link that played a crucial role in the past financial crisis of 2008.

\noindent
 Table \ref{tab:netmes_gelnet} displays mean values of some common network measures to analyze the characteristics of the estimated networks. In fact, network analysis can be a useful tool to study relations among economic entities, see for example Jackson \cite{SEN08} and Borgatti et al. \cite{nass09}. The degree of a node is the number of edges connected to the node. This gives us an indication about how many sectors are connected to one sector. The distance between two nodes is the length of a shortest path joining the two nodes. Ideally the shorter the distance, the faster a shock propagates. Eccentricity of a node is the maximum distance between that node and all other nodes in the network. Eccentricity could give us also a measure of diffusion timing. The (local) clustering coefficient measures the fraction of connections, over the total possible number of connections, among the neighbors of a node. This can suggest us how much the economic sectors tend to cluster together. Lastly, we report the mean value of strength. The strength of a node is the sum of the weights of all edges connected to that node \cite{str04}. Here, the weights are the partial correlations. This measure suggests then how intense, or strong, are overall the connections among sectors and could be used to detect potential crisis period.
 \noindent
 Looking at these characteristics in Table \ref{tab:netmes_gelnet}, we observe a decrease of the mean values of degree and clustering coefficient between 2018 and 2019 followed by an increase between 2019 and 2020. This is in line with the sparser network detected in 2019 (with 23 edges). The average strength follows a similar pattern, with a more noticeable increase between 2019 and 2020 probably fostered by Covid-19 pandemic. The higher mean value of distance in 2019 is also in line with the sparser graph detected and it could also point out a slight increase in time of shock's diffusion. Nonetheless the mean values of distance and eccentricity are small, suggesting that a shock will propagate quickly in the network. Looking at the average eccentricity we can see that, on average, a shock in a sector reach other sectors in a bit more than 2 steps. With only 10 nodes it is difficult to classify these networks in a specific category, nonetheless they show small average distance and noticeable clustering, even if it is not extremely high. These are two core characteristics of small-world graphs \cite{smallworld98}.
 
 \vspace{1cm}
 \begin{table}[H]
    \centering
    \begin{tabular}{l|c|c|c|c}

        \textbf{Measure (mean values)} & \textbf{2018} & \textbf{2019} & \textbf{2020} & \textbf{2018-2020} \\
        \hline
        Degree & 5.400 & 4.600 & 5.400 & 5.200 \\
        \hline
        Distance & 1.422 & 1.511 & 1.422 & 1.422 \\
        \hline
        Eccentricity & 2.200 & 2.200 & 2.200 & 2.000 \\
        \hline
        Clustering & 0.591 & 0.450 & 0.555 & 0.547 \\
        \hline
        Strength & 0.891 & 0.870 & 0.944 & 0.939\\
        
    \end{tabular}
    \caption{Network measures for estimated graphs using 2S-gelnet (AND)}
    \label{tab:netmes_gelnet}
\end{table}

\noindent
Finally, in Table \ref{tab:shocks}, we report results from the simulation of diffusion of a shock. In fact, after estimating the network structure and potential channels of contagion, it is interesting to try to evaluate the potential impact of a shock and the network resilience. Following the method suggested by Anufriev and Panchenko \cite{anuf15}, we use the partial correlation matrix \textbf{P} to simulate a positive shock of magnitude 1 in the financial sector. Then, we rely on equation (\ref{shockdiff}) to evaluate not only the direct effects, but also the second and higher order effects:
\begin{equation} \label{shockdiff}
    \text{\textbf{s}}_{\infty} = \sum_{t=0}^{\infty}\text{\textbf{P}}^t\text{\textbf{e}}=(\text{\textbf{I}}-\text{\textbf{P}})^{-1}\mathbf{e}
\end{equation}
where \textbf{s}$_{\infty}$ is the final steady-state of the effects of the initial vector of shocks \textbf{e}, here \textbf{e} = $[0,0,0,0,1,0,0,0,0,0]^{\top}$, where the fifth element of \textbf{e} corresponds to the financial sector.

 \begin{table}[H]
\centering
\begin{tabular}{l|rrrr|}
& \multicolumn{4}{|c|}{\textbf{Period}} \\
  \hline
 & \textbf{2018} & \textbf{2019} & \textbf{2020} & \textbf{18-20} \\ 
  \hline
Basic materials & 3.207 & 2.643 & 10.745 & 6.356 \\ 
  Consumer cyclicals & 3.502 & 3.039 & 8.229 & 5.582 \\ 
  Consumer non-cyclicals & 1.593 & 1.350 & 6.885 & 3.935 \\ 
  Energy & 1.769 & 1.849 & 6.507 & 4.092 \\ 
  Financials & 3.545 & 3.656 & 12.254 & 7.308 \\ 
  Healthcare & 3.376 & 1.675 & 7.126 & 4.575 \\ 
  Industrials & 4.070 & 3.804 & 12.877 & 8.079 \\ 
  Real estate & 1.093 & 0.679 & 8.674 & 4.412 \\ 
  Technology & 3.502 & 2.981 & 8.062 & 5.653 \\ 
  Utilities & 0.289 & 0.473 & 7.016 & 3.520 \\
   \hline
\end{tabular}
    \caption{Final steady-states after a unitary shock in financial sector}
    \label{tab:shocks}
\end{table}

\begin{figure}[ht]
 \hspace*{-1cm}
 \includegraphics[width=1.1\textwidth]{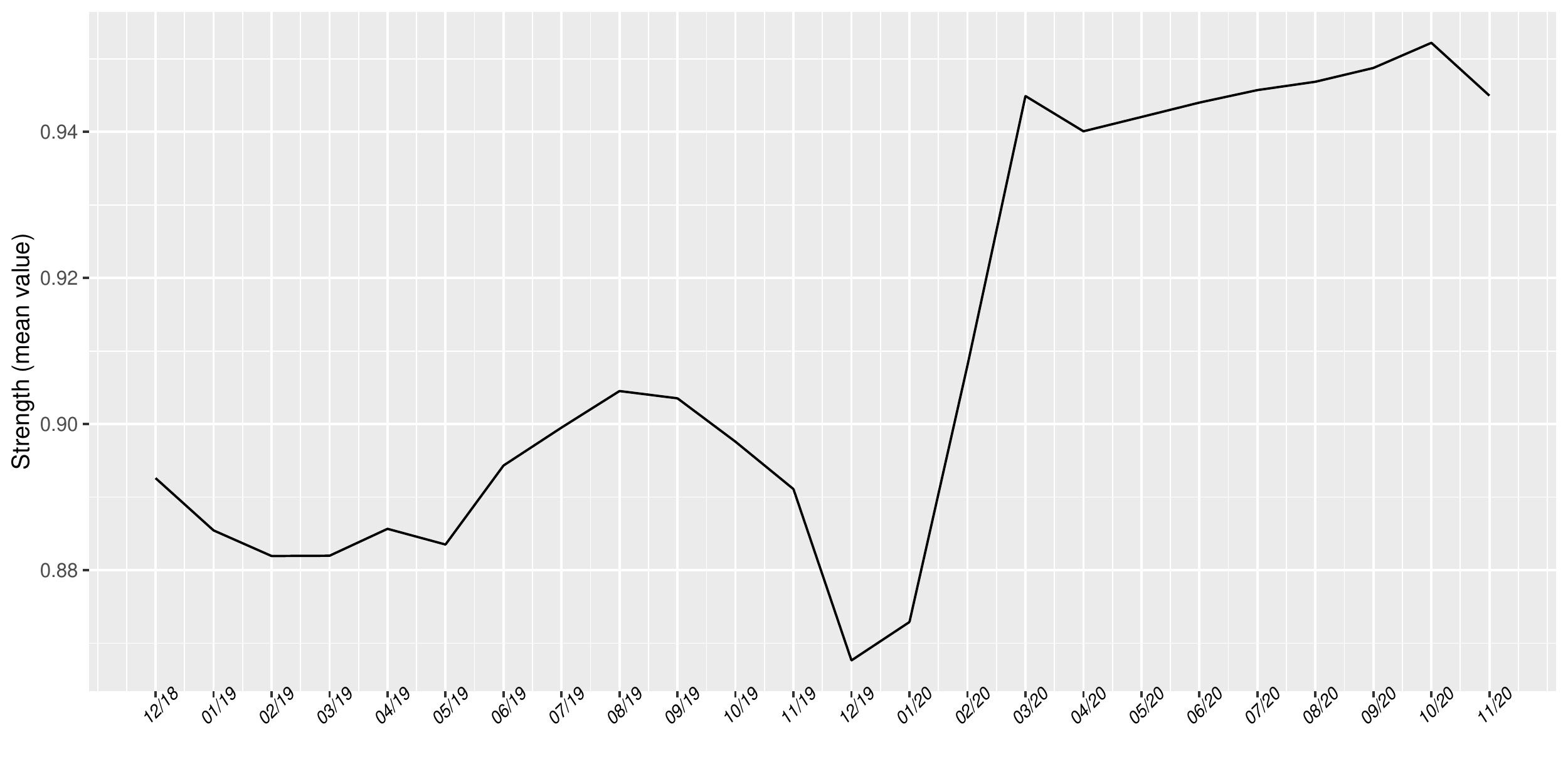}
 \caption{Mean value of strength estimated on a rolling window of 12 months (with shifts of 1 month)}
\label{fig:rollstr}
\end{figure}

\noindent
From Table \ref{tab:shocks} the estimated effect of a shock differs from year to year. So, for example, even if most of the characteristics reported in Table \ref{tab:netmes_gelnet} are similar for 2018 and 2020, the effect of a shock in the financial sector is different in this two years, with a much larger impact in 2020. This could depend not only on a "rewiring" of edges, but also on the strength of the relations among sectors given by partial correlations in \textbf{P}, which can be detected by looking at the weights of the estimated sectoral network (see the mean values of strength in Table \ref{tab:netmes_gelnet}). Figure \ref{fig:rollstr} also plots the evolution of the average strength of networks, estimated using a rolling window approach with window size equal to one year. Notice how the Covid-19 pandemic could have played a role in the increase in average strength, as there is a sharp increase from January 2020 to March 2020. Also the average strength stays at very high level for all 2020, suggesting a prolonged effect on the degree and level of interconnections across sectors, with potential systemic consequences. 

\newpage

\section{Conclusions} \label{sec:conclusions}
\noindent
In this article, we propose three methods to compute a sparse estimate of the precision matrix $\PrecisionN$ in multivariate Gaussian settings. The estimated precision matrix can then be used to reconstruct the conditional dependence graph of the related undirected graphical model. Also, with a proper rescale of $\PrecisionN$, we can derive a symmetric matrix of partial correlations, which can be employed in real-world applications to detect potential dependencies across economic sectors and spillover effects.
All the methods proposed rely on the elastic net penalty. Our first method, \emph{gelnet}, exploits  penalized log-likelihood estimation, while the second method, \emph{CR-gelnet}, consists of $p$ penalized regressions with a subsequent procedure to make the estimate symmetric. The last and third method, \emph{2S-gelnet}, is a two steps procedure that at first estimate the sparsity pattern in the precision matrix and then estimate the precision matrix elements using a log-likelihood procedure with constraints given by the sparsity pattern. We perform extensive simulations to test the proposed methods on a large set of well-known network structures, representing different types of conditional dependence graph in Gaussian graphical models.

\noindent
The main result of our analysis through simulations is that the two stages graphical elastic net (\emph{2S-gelnet}) performs better than the other two, not only accordingly to F$_1$-score and accuracy, but also to Frobenius distance. Using BIC calibration instead of cross-validation is not relevant when we are using \emph{2S-gelnet}, while it leads to different estimated networks when we are using the remaining two methods, \emph{gelnet} and \emph{CR-gelnet}. Moreover, we notice that  the degree of correctness in identification of the network is dependent on the underlying structure. According to our simulations,  the core-periphery topology is in fact the most problematic structure to estimate.

\noindent
Finally, we present an empirical study of the network among 10 US economic sectors. We use the daily prices of sector indices from S\&P500 and the \emph{2S-gelnet} to estimate the relations among these sectors and thereby to estimate the  perceived network. We point out the network connections as well as the properties, evaluating the potential impact of a shock through the  contagion channels. Moreover, we study the evolution of the network strength in time, which increases significantly in correspondence to  the emergence of the Covid-19 pandemic in Europe and USA, suggesting that such an event has potential systemic impact on all the economic sectors.

\section*{Acknowledgments}
The authors would like to thank participants to the \textbf{14$^{\text{th}}$ International Conference on Computational and Financial Econometrics} (December 19-21, 2020) conference for the helpful comments
\newpage
\bibliographystyle{unsrt}
\bibliography{biblio}

\begin{thebibliography}{10}

\bibitem{ElNet05}
H.~Zou and T.~Hastie.
\newblock Regularization and variable selection via the elastic net.
\newblock {\em Journal of the Royal Statistical Society (Series B)},
  67:301--320, 2005.

\bibitem{GM1996}
S.~Lauritzen.
\newblock {\em Graphical Models}.
\newblock Oxford University Press, 1996.

\bibitem{GM2009}
D.~Koller and N.~Friedman.
\newblock {\em Probabilistic Graphical Models: Principles and Techniques}.
\newblock MIT Press, 2009.

\bibitem{anuf15}
M.~Anufriev and V.~Panchenko.
\newblock Connecting the dots: Econometric methods for uncovering networks with
  an application to the australian financial institutions.
\newblock {\em European Journal of Banking and Finance}, 61:241--255, 2015.

\bibitem{Tib96}
R.~Tibshirani.
\newblock Regression shrinkage and selection via the lasso.
\newblock {\em Journal of the Royal Statistical Society. Series B
  (Methodological)}, 58(1):267--288, 1996.

\bibitem{Mein06}
N.~Meinshausen and P.~B{\"u}hlmann.
\newblock High-dimensional graphs and variable selection with the lasso.
\newblock {\em The Annals of Statistics}, 34(3):1436--1462, 2006.

\bibitem{Bane08}
O.~Banerjee, L.~El Ghaoui, and A.~d'Aspremont.
\newblock Model selection through sparse maximum likelihood estimation for
  multivariate gaussian or binary data.
\newblock {\em Journal of Machine Learning Research}, 9:485--516, 2008.

\bibitem{glasso08}
J.~Friedman, T.~Hastie, and R.~Tibshirani.
\newblock Sparse inverse covariance estimation with the graphical lasso.
\newblock {\em Biostatistics}, 9(3):432--441, 2008.

\bibitem{Cucu11}
M.~Cucuringu, J.~Puente, and D.~Shue.
\newblock Model selection in undirected graphical models with the elastic-net.
\newblock {\em arXiv:1111.0559}, 2011.

\bibitem{Ryal12}
S.~Ryali, K.~Supekar T.~Chen, and V.~Menon.
\newblock Estimation of functional connectivity in fmri data using stability
  selection-based sparse partial correlation with elastic net penalty.
\newblock {\em NeuroImage}, 59(4):3852--3861, 2012.

\bibitem{Liu16}
B.~Liu, L.~Jing, J.~Yu, and J.~Li.
\newblock Robust graph learning via constrained elastic-net regularization.
\newblock {\em Neurocomputing}, 171:299--312, 2016.

\bibitem{scalefree99}
A-L. Barabasi and R.~Albert.
\newblock Emergence of scaling in random networks.
\newblock {\em Science}, 286:509--512, 1999.

\bibitem{smallworld98}
D.~J. Watts and S.~H. Strogatz.
\newblock Collective dynamics of 'small-world' networks.
\newblock {\em Nature}, 393:440--442, 1998.

\bibitem{Yuan10}
M.~Yuan.
\newblock High dimensional inverse covariance matrix estimation via linear
  programming.
\newblock {\em Journal of Machine Learning Research}, 11:2261--2286, 2010.

\bibitem{bookCO2004}
S.~Boyd and L.~Vandenberghe.
\newblock {\em Convex Optimization}.
\newblock Cambridge University Press, 2004.

\bibitem{Fried10}
J.~Friedman, T.~Hastie, and R.~Tibshirani.
\newblock Reguralization paths for generalized linear models via coordinate
  descent.
\newblock {\em Journal of Statistical Software}, 33(1):1--22, 2010.

\bibitem{gslope18}
M.~Bogdan, S.~Lee, and P.~Sobczyk.
\newblock Sparse inverse covariance matrix estimation with graphical slope.
\newblock {\em Technical report}, 2018.

\bibitem{bookMA1979}
K.V. Mardia, J.T. Kent, and J.M. Bibby.
\newblock {\em Multivariate Analysis}.
\newblock Academic Press, 1979.

\bibitem{Cai11}
T.~Cai, W.~Liu, and Xi~Luo.
\newblock A constrained l1 minimization approach to sparse precision matrix
  estimation.
\newblock {\em Journal of the American Statistical Association},
  106(494):594--607, 2011.

\bibitem{bookESL2009}
T.~Hastie, R.~Tibshirani, and J.~Friedman.
\newblock {\em The Elements of Statistical Learning: Data Mining, Inference,
  and Prediction}.
\newblock Springer, 2009.

\bibitem{Torri18}
G.~Torri, R.~Giacometti, and S.~Paterlini.
\newblock Robust and sparse banking network estimation.
\newblock {\em European Journal of Operational Research}, 270(1):51--65, 2018.

\bibitem{SHDD11}
P.~B{\"u}hlmann~S. van~de Geer.
\newblock {\em Statistics for High-Dimensional Data: Methods, Theory and
  Applications}.
\newblock Springer, 2011.

\bibitem{ebic10}
R.~Foygel and M.~Drton.
\newblock Extended bayesian information criteria for gaussian graphical models.
\newblock {\em Advances in Neural Information Processing Systems (NIPS 2010)},
  23:604--612, 2010.

\bibitem{ace12}
D.~Acemoglu, V.~M. Carvalho, A.~Ozdaglar, and A.~Tahbaz-Salehi.
\newblock The network origins of aggregate fluctuations.
\newblock {\em Econometrica}, 80:1977--2016, 2012.

\bibitem{SEN08}
M.~O. Jackson.
\newblock {\em Social and Economic Networks}.
\newblock Princeton University Press, 2008.

\bibitem{nass09}
S.~P. Borgatti, A.~Mehra, D.~J. Brass, and G.~Labianca.
\newblock Network analysis in the social sciences.
\newblock {\em Science}, 323:892--895, 2009.

\bibitem{str04}
A.~Barrat, M.~Barthelemy, R.~Pastor-Satorras, and A.~Vespignani.
\newblock The architecture of complex weighted networks.
\newblock {\em PNAS}, 101:3747--3752, 2004.

\end{thebibliography}
\newpage
\begin{appendices}
\section{Adjacency Matrices} \label{app:1}

\begin{figure}[!h]
\includegraphics[width=0.32\textwidth]{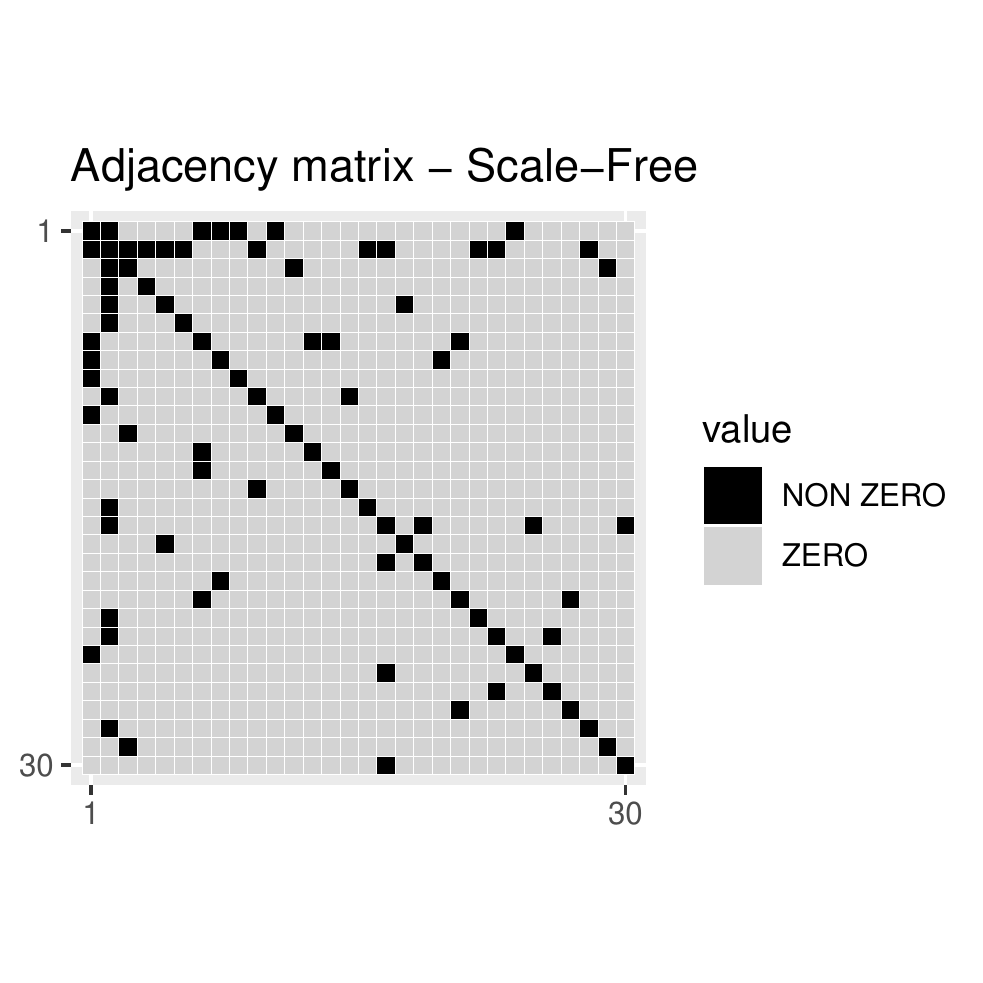}
\includegraphics[width=0.32\textwidth]{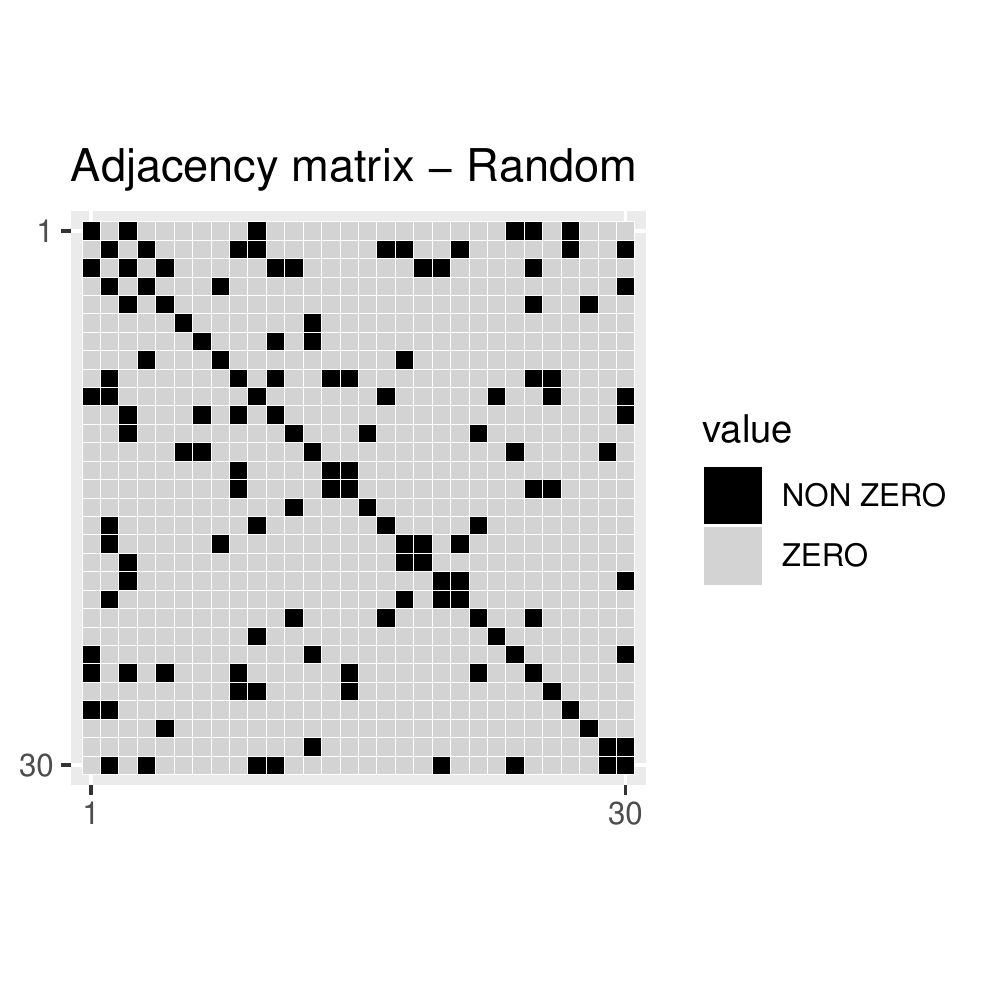}
\includegraphics[width=0.32\textwidth]{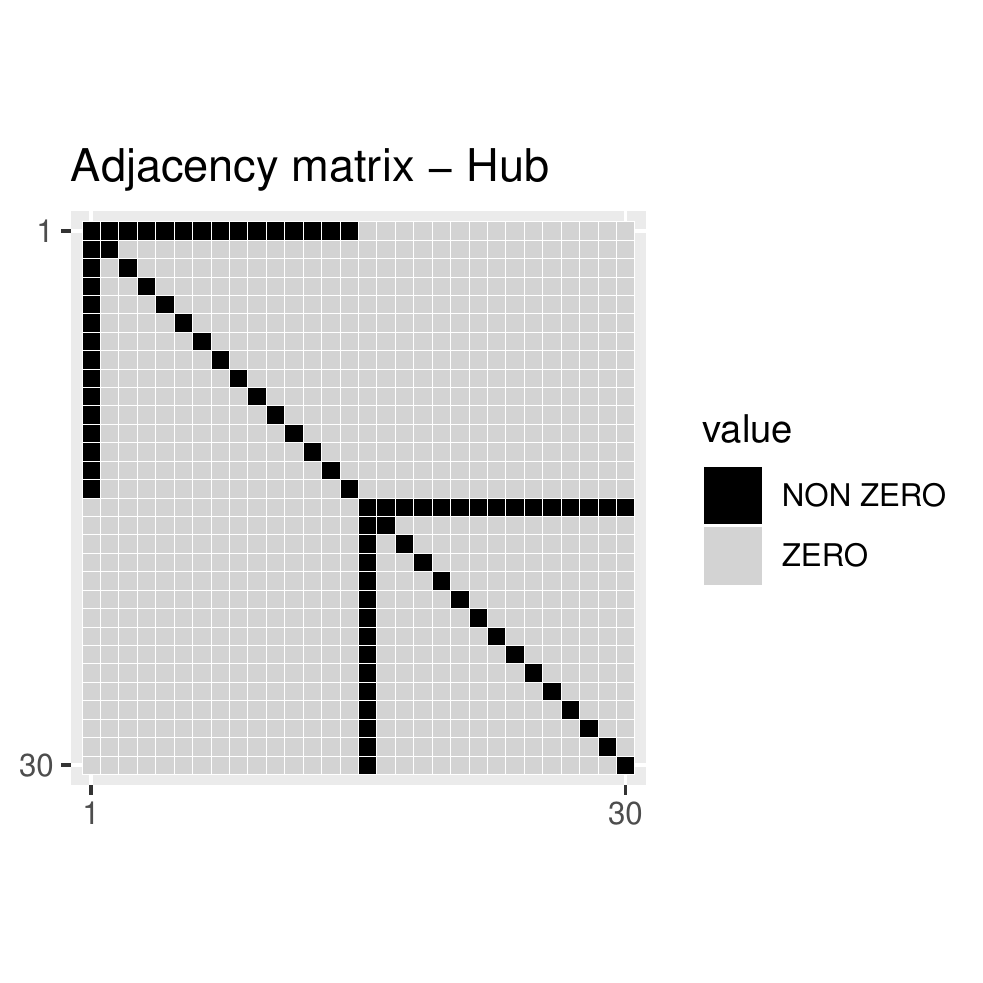}
\end{figure}
\begin{figure}[!h]
\includegraphics[width=0.32\textwidth]{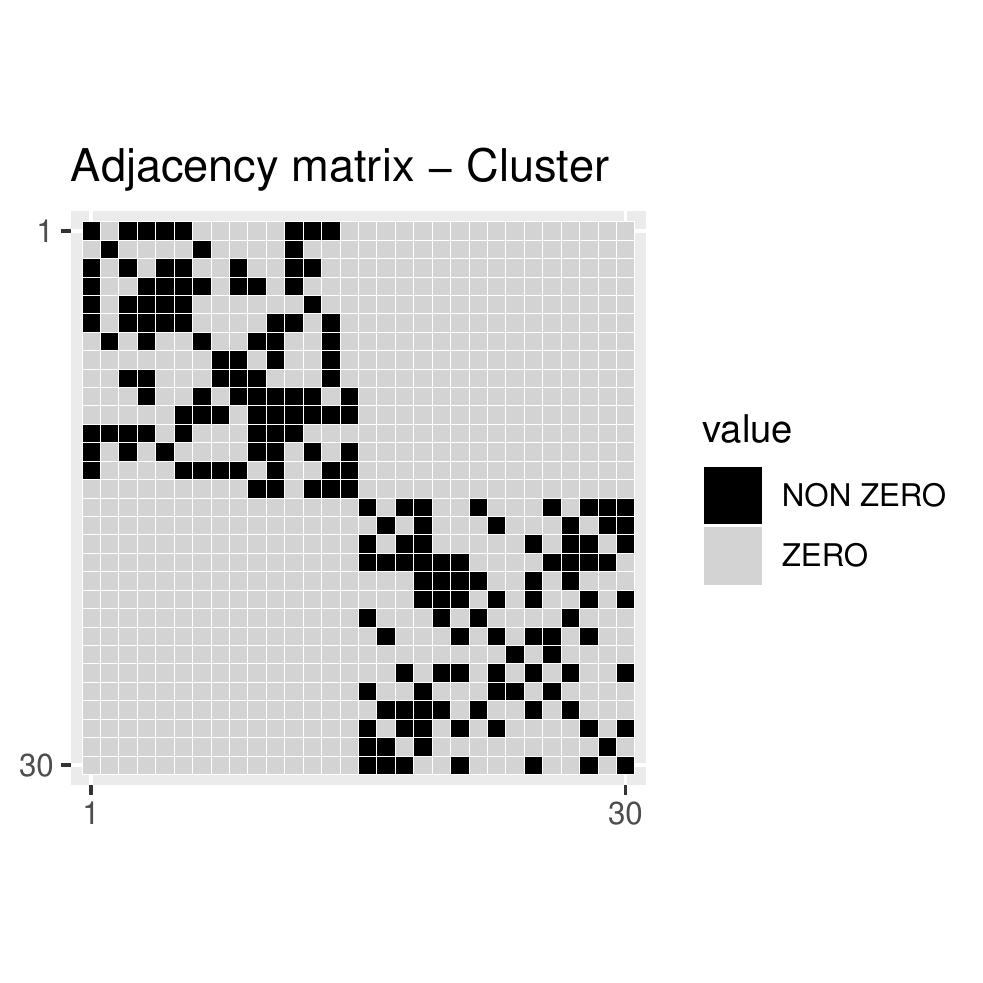}
\includegraphics[width=0.32\textwidth]{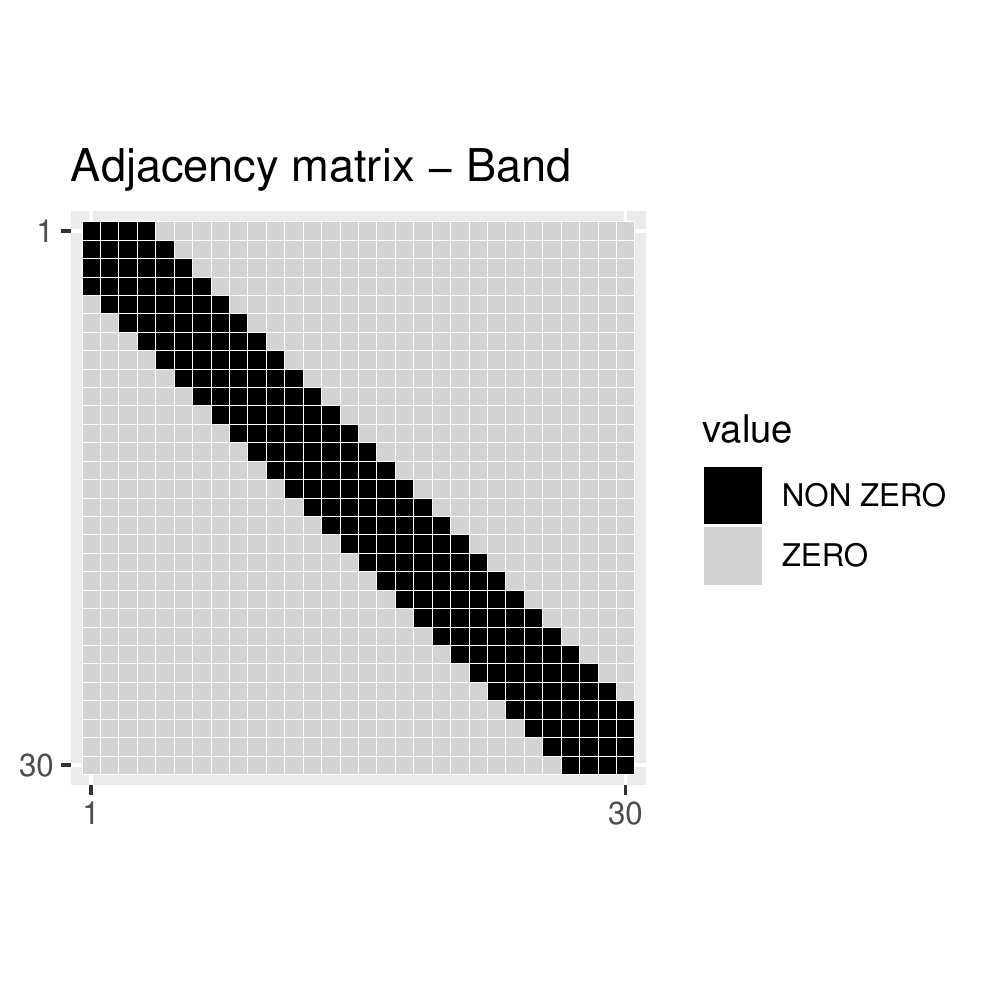}
\includegraphics[width=0.32\textwidth]{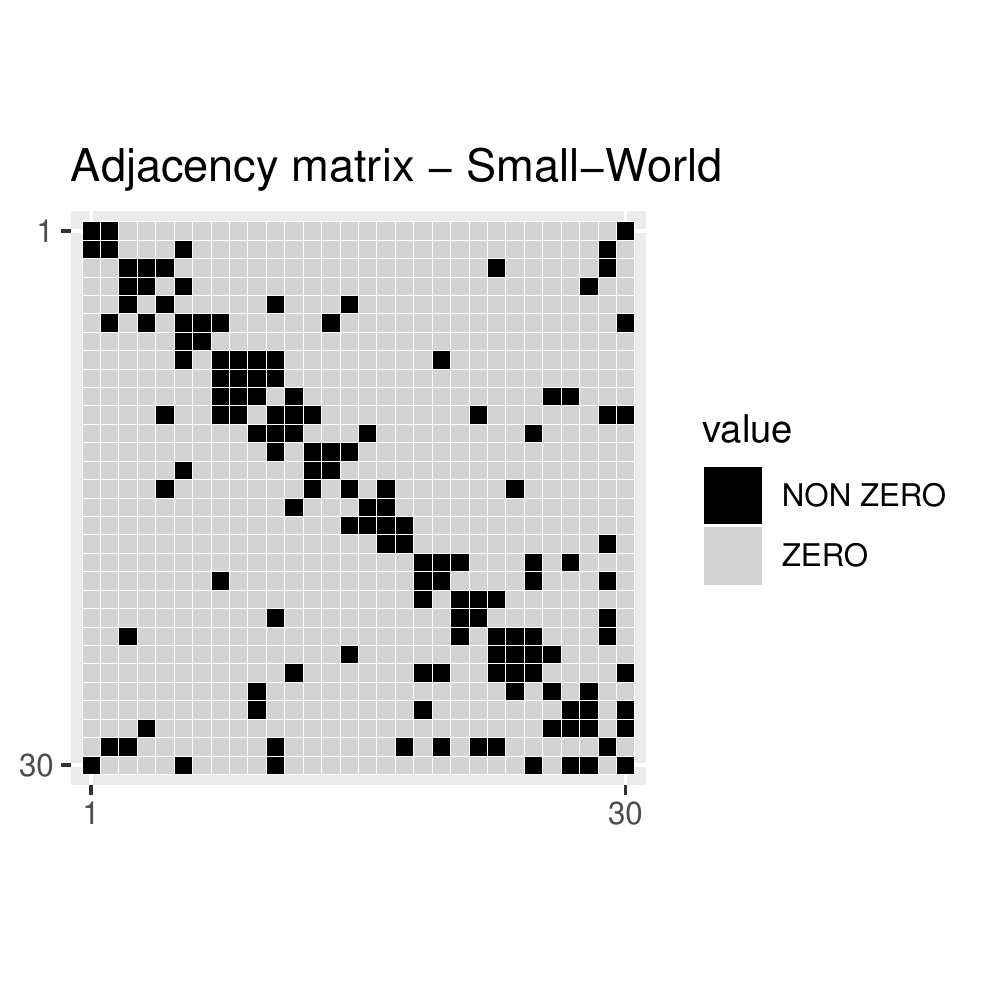}
\end{figure}
\begin{figure}[!h]
\centering
\includegraphics[width=0.32\textwidth]{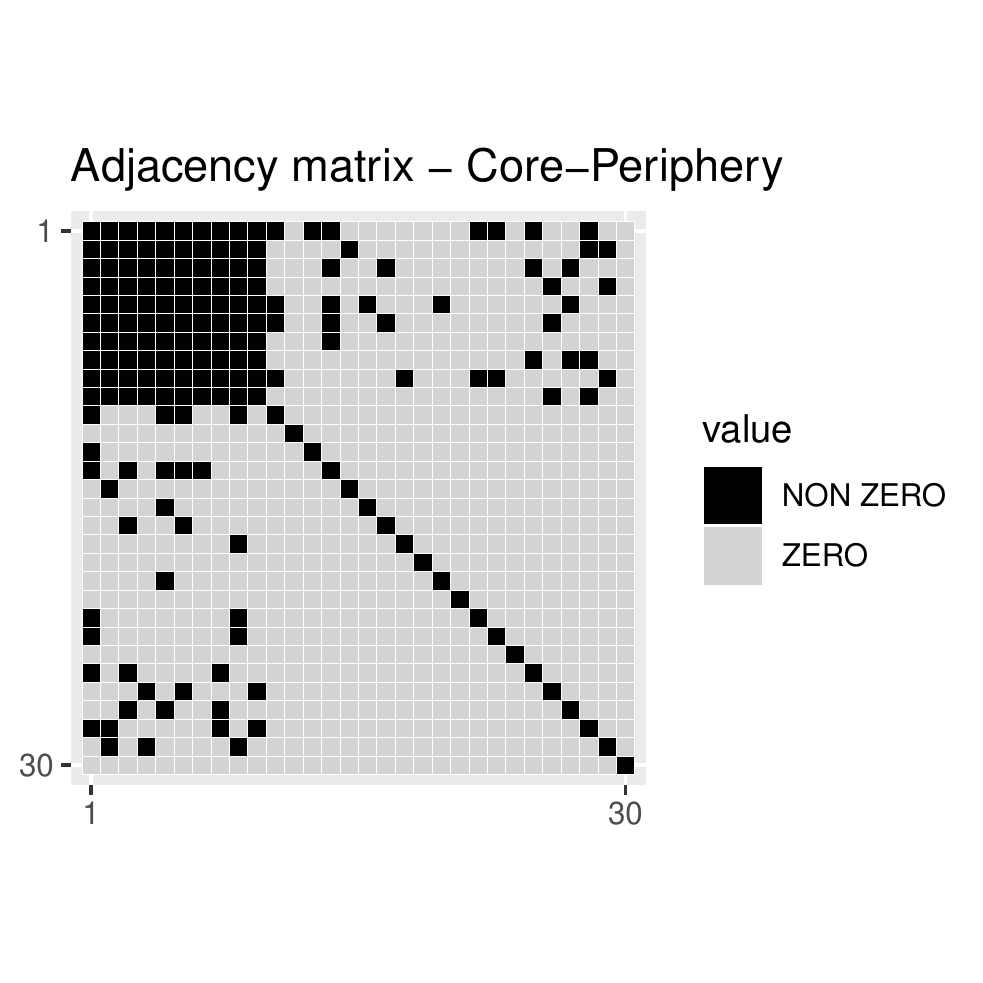}
\end{figure}
\newpage
\section{Additional simulation results}
\subsection{Additional results with sample size = 1000}\label{app:21}

\noindent
Tables \ref{alp1000bic}, \ref{alp1000cv}, \ref{alp200bic} and \ref{alp200cv} show the average value of optimal $\alpha$ (standard deviation in brackets, networks' numbers as listed in Section \ref{sec:setup}, sample size 1000 and 200):
\begin{table}[ht]
\centering
\begin{tabular}{lrrrrrrr}
\hline
                  & Net 1 & Net 2 & Net 3 & Net 4 & Net 5 & Net 6 & Net 7 \\ 
\hline
gelnet            & 0.9992 & 0.9950 & 1.0000 & 0.9967 & 0.9983 & 0.9950 & 0.9917 \\ 
                  & (0.0046) & (0.0121) & (0.0000) & (0.0086) & (0.0063) & (0.0102) & (0.0137) \\ 
2S-gelnet (AND)   & 0.5458 & 0.5383 & 0.4367 & 0.8500 & 0.9150 & 0.6192 & 0.8925 \\ 
                  & (0.2267) & (0.1497) & (0.1685) & (0.0983) & (0.0860) & (0.2418) & (0.0791) \\ 
2S-gelnet (OR)    & 0.5717 & 0.7008 & 0.5133 & 0.9025 & 0.9050 & 0.7133 & 0.9575 \\ 
                  & (0.2444) & (0.2101) & (0.2390) & (0.0925) & (0.1018) & (0.1893) & (0.0451) \\ 
CR-gelnet (L2)    & 0.9883 & 0.9825 & 0.9917 & 0.9742 & 0.9883 & 0.9817 & 0.9767 \\ 
                  & (0.0170) & (0.0199) & (0.0152) & (0.0275) & (0.0205) & (0.0207) & (0.0227) \\ 
CR-gelnet (MinEl) & 0.9817 & 0.9783 & 0.9867 & 0.9750 & 0.9792 & 0.9842 & 0.9700 \\ 
                  & (0.0207) & (0.0260) & (0.0194) & (0.0254) & (0.0294) & (0.0232) & (0.0368) \\ 
\hline
\end{tabular}
\caption{$\alpha$ optimal values using BIC - $n$ = 1000}
\label{alp1000bic}
\end{table}
\begin{table}[!h]
\centering
\begin{tabular}{lrrrrrrr}
\hline
                   & Net 1 & Net 2 & Net 3 & Net 4 & Net 5 & Net 6 & Net 7 \\ 
\hline
gelnet            & 1.0000 & 1.0000 & 1.0000 & 1.0000 & 1.0000 & 1.0000 & 1.0000 \\ 
                  & (0.0000) & (0.0000) & (0.0000) & (0.0000) & (0.0000) & (0.0000) & (0.0000) \\ 
2S-gelnet (AND)   & 0.4358 & 0.7592 & 0.4725 & 0.8508 & 0.8200 & 0.7517 & 0.8642 \\ 
                  & (0.1395) & (0.1902) & (0.0658) & (0.1450) & (0.1833) & (0.1964) & (0.1298) \\ 
2S-gelnet (OR)    & 0.4583 & 0.8158 & 0.4992 & 0.8675 & 0.9258 & 0.8008 & 0.8892 \\ 
                  & (0.1473) & (0.1713) & (0.0860) & (0.1594) & (0.0880) & (0.2024) & (0.0921) \\ 
CR-gelnet (L2)    & 1.0000 & 1.0000 & 0.9992 & 0.9992 & 0.9983 & 1.0000 & 0.9975 \\ 
                  & (0.0000) & (0.0000) & (0.0046) & (0.0046) & (0.0063) & (0.0000) & (0.0076) \\ 
CR-gelnet (MinEl) & 0.9983 & 0.9942 & 1.0000 & 0.9933 & 0.9942 & 0.9992 & 0.9833 \\ 
                  & (0.0063) & (0.0142) & (0.0000) & (0.0245) & (0.0204) & (0.0046) & (0.0379) \\ 
\hline
\end{tabular}
\caption{$\alpha$ optimal values using 5-CV - $n$ = 1000}
\label{alp1000cv}
\end{table}

\newpage
\subsection{Results with sample size = 200}
\label{app:22}

\begin{table}[ht]
\centering
\begin{tabular}{lrrrrrrr}
\hline
                  & Net 1 & Net 2 & Net 3 & Net 4 & Net 5 & Net 6 & Net 7 \\ 
\hline
gelnet            & 0.9992 & 0.9992 & 1.0000 & 0.9958 & 0.9933 & 0.9983 & 0.9975 \\ 
                  & (0.0046) & (0.0046) & (0.0000) & (0.0095) & (0.0130) & (0.0063) & (0.0076) \\ 
2S-gelnet (AND)   & 0.7008 & 0.8108 & 0.7042 & 0.9225 & 0.9208 & 0.7808 & 0.7867 \\ 
                  & (0.2121) & (0.1469) & (0.1934) & (0.0752) & (0.0763) & (0.1700) & (0.1750) \\ 
2S-gelnet (OR)    & 0.7325 & 0.8267 & 0.7508 & 0.9208 & 0.9617 & 0.8558 & 0.8158 \\ 
                  & (0.1786) & (0.1129) & (0.1613) & (0.0838) & (0.0439) & (0.1471) & (0.1743) \\ 
CR-gelnet (L2)    & 0.9925 & 0.9858 & 0.9950 & 0.9858 & 0.9908 & 0.9867 & 0.9908 \\ 
                  & (0.0238) & (0.0224) & (0.0121) & (0.0276) & (0.0154) & (0.0170) & (0.0167) \\ 
CR-gelnet (MinEl) & 0.9900 & 0.9767 & 0.9842 & 0.9833 & 0.9800 & 0.9817 & 0.9808 \\ 
                  & (0.0155) & (0.0270) & (0.0180) & (0.0240) & (0.0240) & (0.0245) & (0.0252) \\ 
\hline
\end{tabular}
\caption{$\alpha$ optimal values using BIC - $n$ = 200}
\label{alp200bic}
\end{table}

\begin{table}[ht]
\centering
\begin{tabular}{lrrrrrrr}
\hline
                  & Net 1 & Net 2 & Net 3 & Net 4 & Net 5 & Net 6 & Net 7 \\ 
\hline
gelnet            & 1.0000 & 1.0000 & 1.0000 & 1.0000 & 1.0000 & 1.0000 & 0.9358 \\ 
                  & (0.0000) & (0.0000) & (0.0000) & (0.0000) & (0.0000) & (0.0000) & (0.1324) \\ 
2S-gelnet (AND)   & 0.5958 & 0.6242 & 0.5483 & 0.7825 & 0.9067 & 0.6400 & 0.5025 \\ 
                  & (0.1662) & (0.2238) & (0.1393) & (0.1930) & (0.1006) & (0.2165) & (0.1413) \\ 
2S-gelnet (OR)    & 0.8150 & 0.7517 & 0.8292 & 0.7267 & 0.9008 & 0.7642 & 0.6892 \\ 
                  & (0.1511) & (0.2065) & (0.1292) & (0.2093) & (0.1146) & (0.1954) & (0.1946) \\ 
CR-gelnet (L2)    & 1.0000 & 1.0000 & 1.0000 & 0.9992 & 1.0000 & 1.0000 & 0.9983 \\ 
                  & (0.0000) & (0.0000) & (0.0000) & (0.0046) & (0.0000) & (0.0000) & (0.0063) \\ 
CR-gelnet (MinEl) & 0.9900 & 0.9992 & 1.0000 & 0.9958 & 1.0000 & 0.9975 & 0.5392 \\ 
                  & (0.0548) & (0.0046) & (0.0000) & (0.0115) & (0.0000) & (0.0076) & (0.3234) \\ 
\hline
\end{tabular}
\caption{$\alpha$ optimal values using 5-CV - $n$ = 200}
\label{alp200cv}
\end{table}

\clearpage

\begin{table}[ht]
\centering
\begin{tabular}{rrrrrrr}
  \hline
 & gelnet & 2S-gelnet & 2S-gelnet & CR-gelnet & CR-gelnet & glasso \\
 &  & (AND) & (OR) & (L2) & (MinEl) &  \\
  \hline
Scale-Free       & 0.9472 & 0.9754 & \textbf{0.9823} & 0.9550 & 0.9530 & 0.9470 \\ 
                 & (0.0195) & (0.0094) & (0.0065) & (0.0171) & (0.0139) & (0.0195) \\ 
Random           & 0.9005 & 0.9677 & \textbf{0.9695} & 0.9172 & 0.9231 & 0.9004 \\ 
                 & (0.0262) & (0.0134) & (0.0081) & (0.0285) & (0.0231) & (0.0266) \\ 
Hub              & 0.9217 & 0.9762 & \textbf{0.9892} & 0.9436 & 0.9349 & 0.9217 \\ 
                 & (0.0267) & (0.0100) & (0.0076) & (0.0204) & (0.0274) & (0.0267) \\ 
Cluster          & 0.8574 & \textbf{0.9312} & 0.9187 & 0.8669 & 0.8773 & 0.8569 \\ 
                 & (0.0319) & (0.0114) & (0.0133) & (0.0285) & (0.0314) & (0.0325) \\ 
Band             & 0.8031 & \textbf{0.9178} & 0.8979 & 0.8360 & 0.8480 & 0.8031 \\ 
                 & (0.0390) & (0.0148) & (0.0158) & (0.0470) & (0.0370) & (0.0398) \\ 
Small-World      & 0.8828 & \textbf{0.9691} & 0.9638 & 0.9100 & 0.9112 & 0.8828 \\ 
                 & (0.0254) & (0.0089) & (0.0103) & (0.0238) & (0.0284) & (0.0256) \\ 
Core-Periphery   & 0.8397 & 0.8734 & \textbf{0.8787} & 0.8574 & 0.8658 & 0.8405 \\ 
                 & (0.0223) & (0.0098) & (0.0098) & (0.0173) & (0.0154) & (0.0214) \\
   \hline
\end{tabular}
\caption{Average Accuracy (Std. Dev. in brackets) - BIC calibration ($n$ = 200)}
\label{acc200bic}
\end{table}

\begin{table}[!h]
\centering
\begin{tabular}{rrrrrrr}
  \hline
 & gelnet & 2S-gelnet & 2S-gelnet & CR-gelnet & CR-gelnet & glasso \\
 &  & (AND) & (OR) & (L2) & (MinEl) &  \\
  \hline
Scale-Free       & 0.8218 & 0.9785 & \textbf{0.9841} & 0.8311 & 0.8631 & 0.8283 \\ 
                 & (0.0342) & (0.0081) & (0.0066) & (0.0331) & (0.0315) & (0.0337) \\ 
Random           & 0.7398 & 0.9641 & \textbf{0.9664} & 0.7513 & 0.7990 & 0.7484 \\ 
                 & (0.0366) & (0.0162) & (0.0135) & (0.0309) & (0.0374) & (0.0332) \\ 
Hub              & 0.7926 & 0.9711 & \textbf{0.9942} & 0.7804 & 0.8215 & 0.7958 \\ 
                 & (0.0266) & (0.0193) & (0.0072) & (0.0335) & (0.0402) & (0.0275) \\ 
Cluster          & 0.6702 & \textbf{0.9149} & 0.9018 & 0.6898 & 0.7289 & 0.6789 \\ 
                 & (0.0321) & (0.0184) & (0.0206) & (0.0263) & (0.0405) & (0.0342) \\ 
Band             & 0.5763 & \textbf{0.8966} & 0.8617 & 0.6381 & 0.7078 & 0.5895 \\ 
                 & (0.0357) & (0.0224) & (0.0296) & (0.0419) & (0.0563) & (0.0392) \\ 
Small-World      & 0.7520 & \textbf{0.9589} & 0.9581 & 0.7480 & 0.7918 & 0.7566 \\ 
                 & (0.0306) & (0.0152) & (0.0141) & (0.0360) & (0.0432) & (0.0281) \\ 
Core-Periphery   & 0.6588 & 0.8759 & \textbf{0.8771} & 0.7137 & 0.6870 & 0.6710 \\ 
                 & (0.0413) & (0.0127) & (0.0102) & (0.0325) & (0.0665) & (0.0312) \\ 
   \hline
\end{tabular}
\caption{Average Accuracy (Std. Dev. in brackets) - 5-CV calibration ($n$ = 200)}
\label{acc200cv}
\end{table}

\newpage

\begin{table}[ht]
\centering
\begin{tabular}{rrrrrrr}
  \hline
 & gelnet & 2S-gelnet & 2S-gelnet & CR-gelnet & CR-gelnet & glasso \\
 &  & (AND) & (OR) & (L2) & (MinEl) &  \\ 
  \hline
Scale-Free       & 0.7162 & 0.8390 & \textbf{0.8783} & 0.7500 & 0.7358 & 0.7153 \\ 
                 & (0.0726) & (0.0520) & (0.0416) & (0.0676) & (0.0587) & (0.0728) \\ 
Random           & 0.7041 & 0.8774 & \textbf{0.8822} & 0.7443 & 0.7560 & 0.7039 \\ 
                 & (0.0551) & (0.0430) & (0.0275) & (0.0660) & (0.0555) & (0.0556) \\ 
Hub              & 0.6316 & 0.8417 & \textbf{0.9247} & 0.7022 & 0.6702 & 0.6316 \\ 
                 & (0.0825) & (0.0590) & (0.0484) & (0.0701) & (0.0937) & (0.0825) \\ 
Cluster          & 0.7006 & \textbf{0.8301} & 0.8035 & 0.7301 & 0.7505 & 0.7003 \\ 
                 & (0.0374) & (0.0268) & (0.0272) & (0.0371) & (0.0444) & (0.0382) \\ 
Band             & 0.6267 & \textbf{0.8178} & 0.7759 & 0.7016 & 0.7197 & 0.6268 \\ 
                 & (0.0330) & (0.0272) & (0.0286) & (0.0541) & (0.0482) & (0.0339) \\ 
Small-World      & 0.6915 & \textbf{0.8900} & 0.8730 & 0.7468 & 0.7516 & 0.6914 \\ 
                 & (0.0437) & (0.0293) & (0.0331) & (0.0490) & (0.0600) & (0.0443) \\ 
Core-Periphery   & 0.5878 & 0.6118 & \textbf{0.6239} & 0.6122 & 0.6066 & 0.5888 \\ 
                 & (0.0267) & (0.0253) & (0.0320) & (0.0265) & (0.0281) & (0.0257) \\ 
   \hline
\end{tabular}
\caption{Average F$_1$-score (Std. Dev. in brackets) - BIC calibration ($n$ = 200)}
\label{f1m200bic}
\end{table}

\begin{table}[!h]
\centering
\begin{tabular}{rrrrrrr}
  \hline
 & gelnet & 2S-gelnet & 2S-gelnet & CR-gelnet & CR-gelnet & glasso \\
 &  & (AND) & (OR) & (L2) & (MinEl) &  \\ 
  \hline
Scale-Free       & 0.4319 & 0.8492 & \textbf{0.8834} & 0.4451 & 0.4968 & 0.4410 \\ 
                 & (0.0454) & (0.0468) & (0.0449) & (0.0483) & (0.0569) & (0.0454) \\ 
Random           & 0.4800 & 0.8623 & \textbf{0.8691} & 0.4915 & 0.5456 & 0.4881 \\ 
                 & (0.0341) & (0.0500) & (0.0392) & (0.0303) & (0.0455) & (0.0323) \\ 
Hub              & 0.3854 & 0.8184 & \textbf{0.9587} & 0.3725 & 0.4241 & 0.3891 \\ 
                 & (0.0314) & (0.0890) & (0.0466) & (0.0349) & (0.0603) & (0.0315) \\ 
Cluster          & 0.5326 & \textbf{0.7992} & 0.7727 & 0.5479 & 0.5824 & 0.5391 \\ 
                 & (0.0234) & (0.0330) & (0.0319) & (0.0204) & (0.0347) & (0.0261) \\ 
Band             & 0.4776 & \textbf{0.7846} & 0.7301 & 0.5177 & 0.5731 & 0.4856 \\ 
                 & (0.0212) & (0.0341) & (0.0379) & (0.0286) & (0.0471) & (0.0238) \\ 
Small-World      & 0.5191 & \textbf{0.8599} & 0.8566 & 0.5157 & 0.5649 & 0.5234 \\ 
                 & (0.0307) & (0.0422) & (0.0402) & (0.0366) & (0.0510) & (0.0292) \\ 
Core-Periphery   & 0.4691 & 0.5963 & \textbf{0.6140} & 0.5168 & 0.5102 & 0.4752 \\ 
                 & (0.0281) & (0.0304) & (0.0290) & (0.0247) & (0.0412) & (0.0234) \\ 
   \hline
\end{tabular}
\caption{Average F$_1$-score (Std. Dev. in brackets) - 5-CV calibration ($n$ = 200)}
\label{f1m200cv}
\end{table}

\clearpage

\begin{table}[ht]
\centering
\begin{tabular}{rrrrrrr}
  \hline
 & gelnet & 2S-gelnet & 2S-gelnet & CR-gelnet & CR-gelnet & glasso \\
 &  & (AND) & (OR) & (L2) & (MinEl) &  \\
  \hline
Scale-Free       & 0.9606 & 0.8646 & \textbf{0.7754} & 1.0110 & 1.0137 & 0.9600 \\ 
                 & (0.0855) & (0.1244) & (0.1220) & (0.0903) & (0.0847) & (0.0856) \\ 
Random           & 1.3067 & 0.9879 & \textbf{0.9766} & 1.3110 & 1.2653 & 1.3064 \\ 
                 & (0.0819) & (0.1040) & (0.0972) & (0.1025) & (0.1011) & (0.0825) \\ 
Hub              & 0.6730 & 0.8162 & \textbf{0.5948} & 0.7923 & 0.8792 & 0.6730 \\ 
                 & (0.0694) & (0.1600) & (0.1534) & (0.0636) & (0.0662) & (0.0694) \\ 
Cluster          & 2.0306 & \textbf{1.4731} & 1.5541 & 1.8228 & 1.6742 & 2.0291 \\ 
                 & (0.1764) & (0.1645) & (0.1714) & (0.1710) & (0.1420) & (0.1768) \\ 
Band             & 2.7983 & \textbf{1.4456} & 1.7339 & 2.2333 & 1.8659 & 2.7995 \\ 
                 & (0.2547) & (0.2668) & (0.3060) & (0.2748) & (0.1753) & (0.2553) \\ 
Small-World      & 1.2784 & \textbf{1.0332} & 1.0564 & 1.3079 & 1.2450 & 1.2789 \\ 
                 & (0.1155) & (0.1243) & (0.1312) & (0.1160) & (0.1000) & (0.1153) \\ 
Core-Periphery   & 2.6523 & 2.5540 & \textbf{2.5386} & 2.6623 & 2.6952 & 2.6535 \\ 
                 & (0.0521) & (0.0658) & (0.0766) & (0.0718) & (0.0586) & (0.0509) \\ 
   \hline
\end{tabular}
\caption{Average Frobenius distance (Std. Dev. in brackets) - BIC calibration ($n$ = 200)}
\label{fd200bic}
\end{table}

\begin{table}[!h]
\centering
\begin{tabular}{rrrrrrr}
  \hline
 & gelnet & 2S-gelnet & 2S-gelnet & CR-gelnet & CR-gelnet & glasso \\
 &  & (AND) & (OR) & (L2) & (MinEl) &  \\
  \hline
Scale-Free       & 0.8710 & 0.8818 & \textbf{0.7911} & 0.8881 & 0.9215 & 0.8684 \\ 
                 & (0.0527) & (0.0987) & (0.1309) & (0.0554) & (0.0587) & (0.0522) \\ 
Random           & 1.1257 & 1.0743 & \textbf{1.0670} & 1.1049 & 1.1260 & 1.1262 \\ 
                 & (0.0605) & (0.1381) & (0.1175) & (0.0661) & (0.0695) & (0.0596) \\ 
Hub              & 0.8164 & 0.8538 & \textbf{0.4724} & 0.8432 & 0.8944 & 0.8139 \\ 
                 & (0.0800) & (0.1698) & (0.1922) & (0.0841) & (0.0848) & (0.0804) \\ 
Cluster          & 1.5540 & 1.5160 & 1.5786 & 1.4468 & \textbf{1.4244} & 1.5650 \\ 
                 & (0.0841) & (0.2032) & (0.2450) & (0.0997) & (0.0893) & (0.0870) \\ 
Band             & 1.8879 & \textbf{1.4020} & 1.5212 & 1.6073 & 1.5173 & 1.9210 \\ 
                 & (0.1100) & (0.2447) & (0.2680) & (0.1080) & (0.1061) & (0.1265) \\ 
Small-World      & 1.1340 & 1.0784 & \textbf{1.0780} & 1.1288 & 1.1505 & 1.1338 \\ 
                 & (0.0780) & (0.1346) & (0.1328) & (0.0809) & (0.0854) & (0.0783) \\ 
Core-Periphery   & 2.4513 & 2.5927 & 2.5544 & 2.3939 & \textbf{2.3299} & 2.4544 \\ 
                 & (0.0539) & (0.0683) & (0.0671) & (0.0623) & (0.1323) & (0.0463) \\ 
   \hline
\end{tabular}
\caption{Average Frobenius distance (Std. Dev. in brackets) - 5-CV calibration ($n$ = 200)}
\label{fd200cv}
\end{table}

\clearpage

\begin{figure}[h!]
\begin{subfigure}{1\textwidth}
    \hspace{-1cm}
  \includegraphics[width=1.2\linewidth]{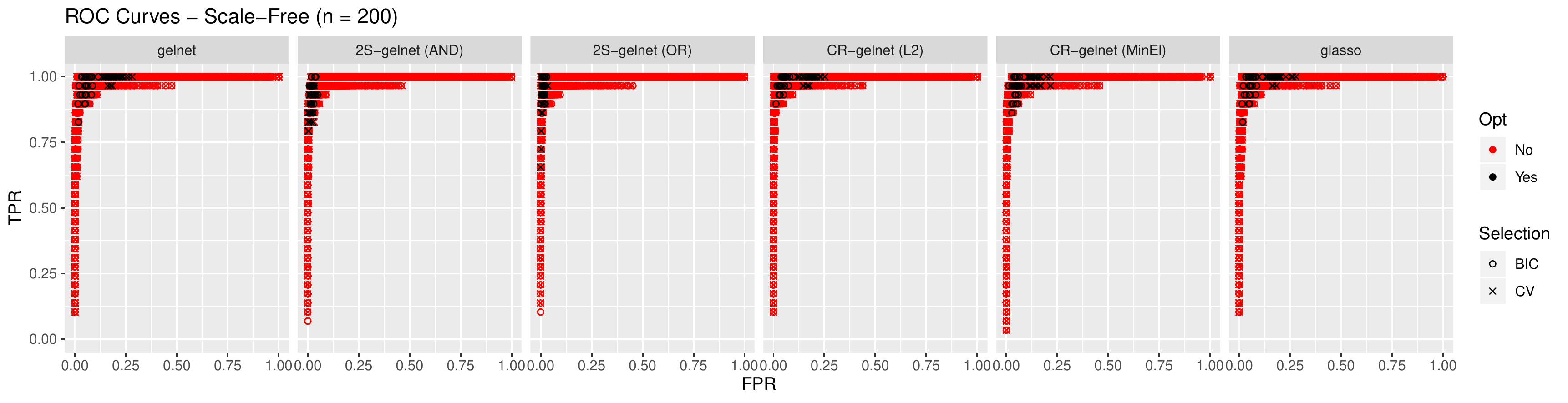}
\end{subfigure}%
\newline
\begin{subfigure}{1\textwidth}
\hspace{-1cm}
  \includegraphics[width=1.2\linewidth]{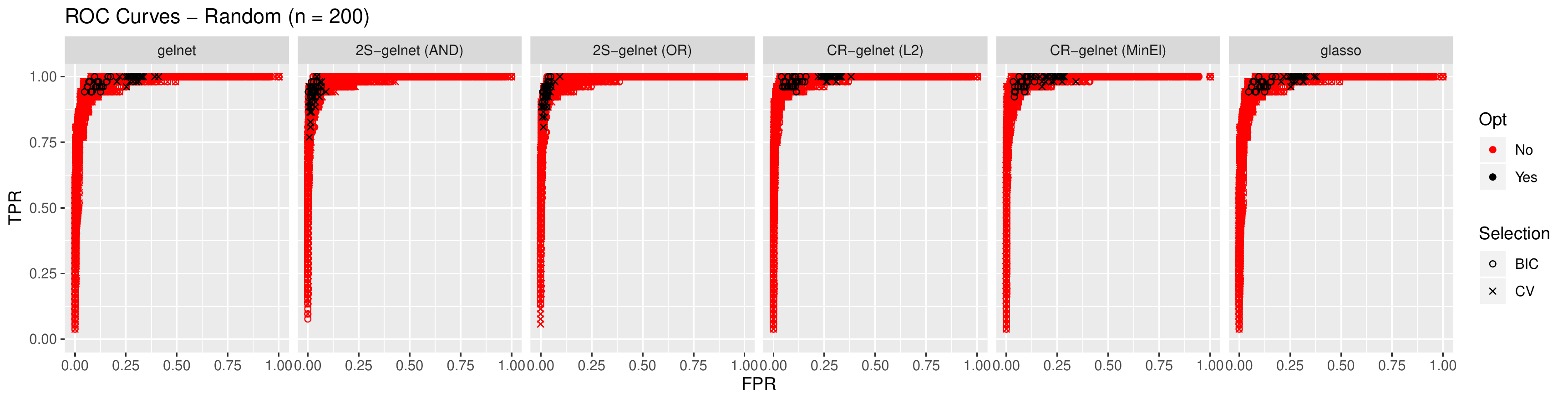}
\end{subfigure}
\newline
\begin{subfigure}{1\textwidth}
\hspace{-1cm}
  \includegraphics[width=1.2\linewidth]{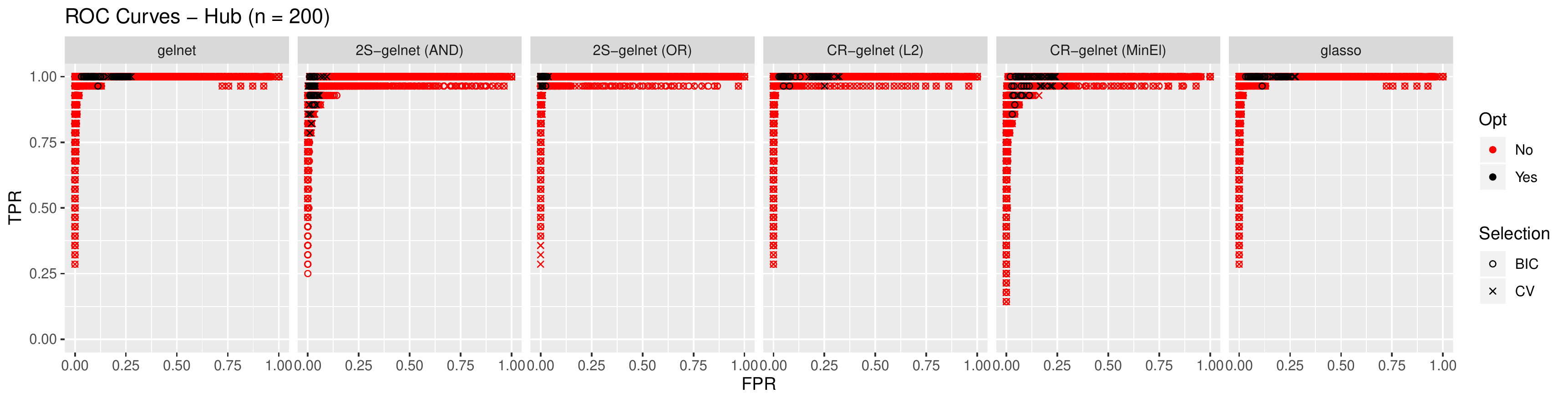}
\end{subfigure}
\begin{subfigure}{1\textwidth}
\hspace{-1cm}
  \includegraphics[width=1.2\linewidth]{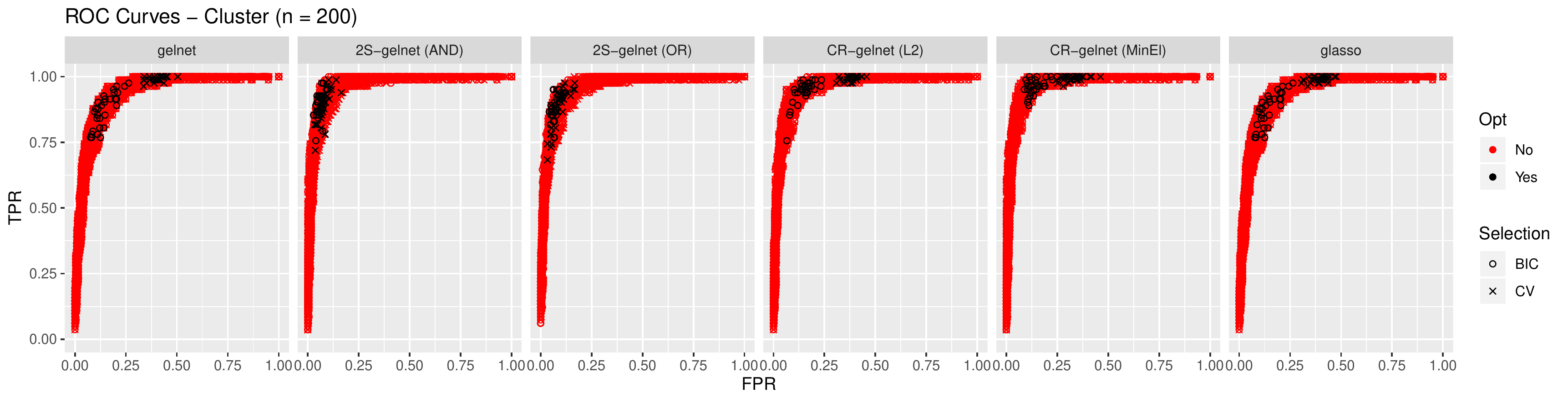}
\end{subfigure}
\end{figure}
\clearpage
\begin{figure}[ht]\ContinuedFloat
\begin{subfigure}{1\textwidth}
\hspace{-1cm}
  \includegraphics[width=1.2\linewidth]{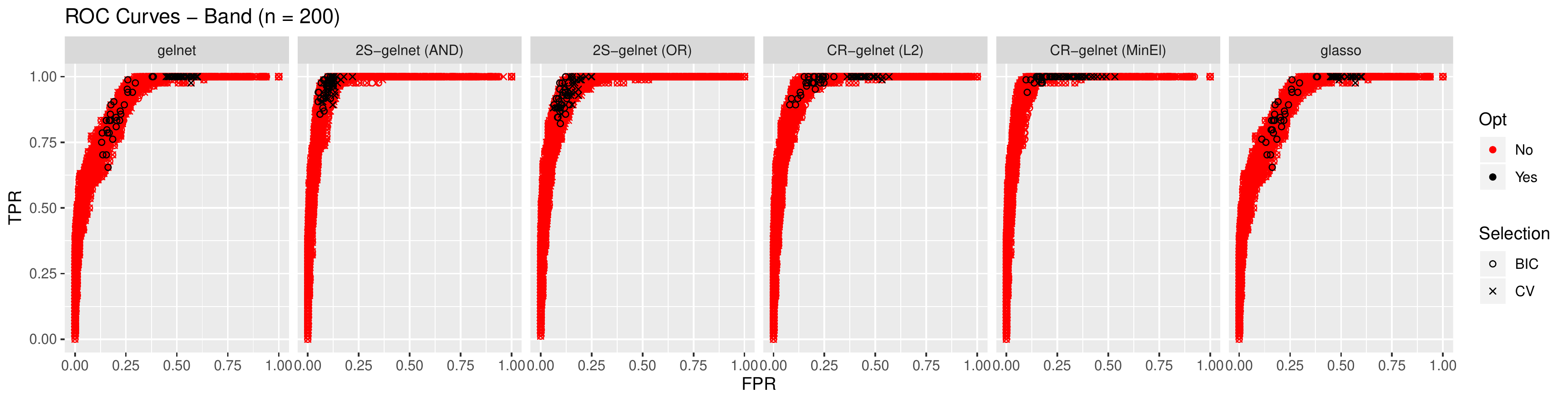}
\end{subfigure}
\newline
\begin{subfigure}{1\textwidth}
\hspace{-1cm}
  \includegraphics[width=1.2\linewidth]{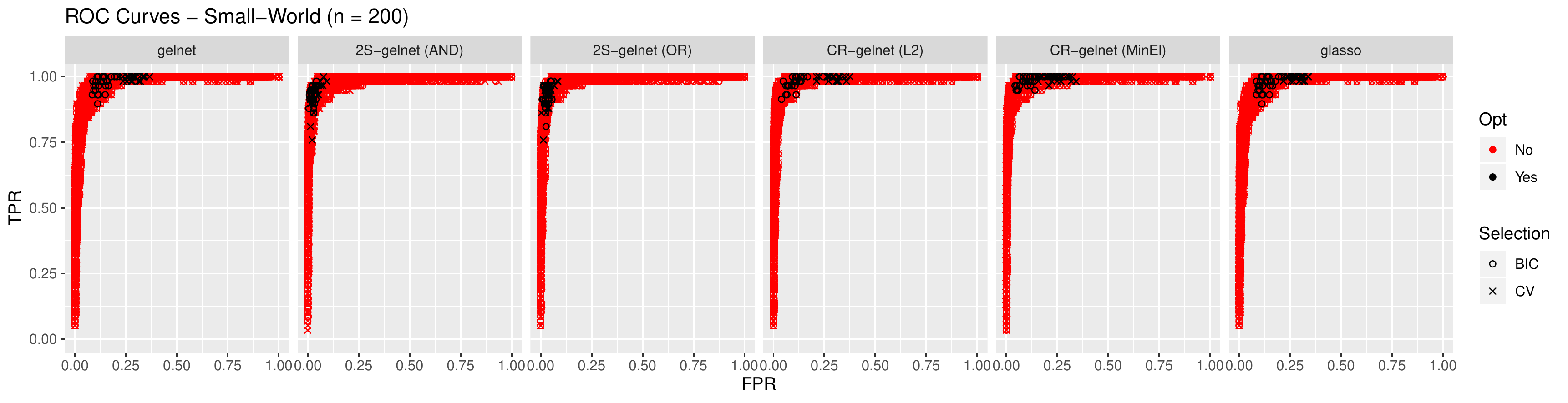}
\end{subfigure}
\newline
\begin{subfigure}{1\textwidth}
\hspace{-1cm}
  \includegraphics[width=1.2\linewidth]{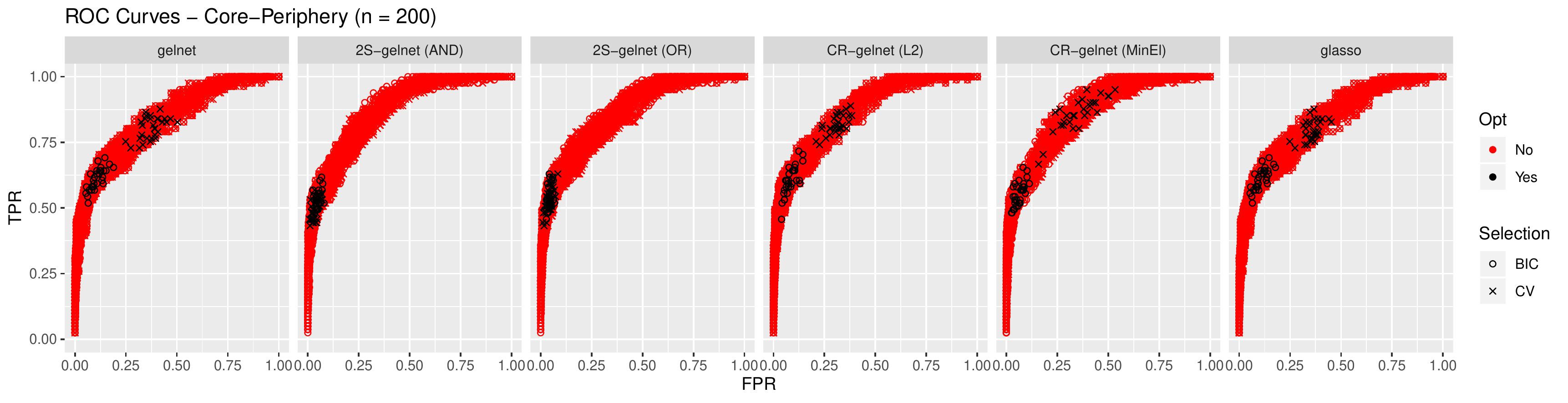}
\end{subfigure}
\caption{ROC curves for seven network structures (in rows) for gelnet (column 1), 2S-gelnet (columns 2 \& 3), CR-gelnet (columns 4 \& 5) and glasso (column 6) - (n=200)}
\label{fig:roc200}
\end{figure}
\end{appendices}

\end{document}